\documentclass[twocolumn,trackchanges]{aastex63}
\usepackage{booktabs}
\usepackage[flushleft]{threeparttable}

\received{Soon}
\revised{After that}
\accepted{\today}

\submitjournal{ApJ}

\shorttitle{Strong NIR Carbon in SN~2015bp}
\shortauthors{Wyatt et al.}

\begin{document}

\title{Strong Near-Infrared Carbon Absorption in the Transitional Type Ia SN~2015bp \footnote{This paper includes data gathered with the 6.5\,m Magellan Telescope at Las Campanas Observatory, Chile.}}

\correspondingauthor{Samuel D. Wyatt}
\email{swyatt@email.arizona.edu}

\author[0000-0003-2732-4956]{S. D. Wyatt}
\affiliation{Steward Observatory, University of Arizona, 933 North Cherry Avenue, Rm. N204, Tucson, AZ 85721-0065, USA}

\author[0000-0003-4102-380X]{D. J. Sand}
\affiliation{Steward Observatory, University of Arizona, 933 North Cherry Avenue, Rm. N204, Tucson, AZ 85721-0065, USA}

\author[0000-0003-1039-2928]{E. Y. Hsiao}
\affil{Department of Physics, Florida State University, Tallahassee, FL 32306, USA}

\author[0000-0003-4625-6629]{C. R. Burns}
\affil{Observatories of the Carnegie Institution for Science, 813 Santa Barbara St, Pasadena, CA 91101, USA}

\author[0000-0001-8818-0795]{S. Valenti}
\affil{Department of Physics, University of California, 1 Shields Avenue, Davis, CA 95616-5270, USA}

\author[0000-0002-4924-444X]{K. A. Bostroem}
\affil{Department of Physics, University of California, 1 Shields Avenue, Davis, CA 95616-5270, USA}

\author[0000-0001-9589-3793]{M. Lundquist}
\affiliation{Steward Observatory, University of Arizona, 933 North Cherry Avenue, Rm. N204, Tucson, AZ 85721-0065, USA}

\author[0000-0002-1296-6887]{L. Galbany}
\affil{Departamento de Fısica Teorica y del Cosmos, Universidad de Granada, E-18071 Granada, Spain}

\author[0000-0002-3900-1452]{J. Lu}
\affil{Department of Physics, Florida State University, Tallahassee, FL 32306, USA}

\author[0000-0002-5221-7557]{C. Ashall}
\affil{Department of Physics, Florida State University, Tallahassee, FL 32306, USA}

\author[0000-0002-0805-1908]{T. R. Diamond}
\affil{Private Astronomer: tiaradiamond@gmail.com}

\author[0000-0003-3460-0103]{A. V. Filippenko}
\affil{Department of Astronomy, University of California, Berkeley, CA 94720-3411, USA}
\affil{Miller Senior Fellow, Miller Institute for Basic Research in Science, University of California, Berkeley, CA 94720, USA}

\author[0000-0002-9154-3136]{M. L. Graham}
\affil{Department of Astronomy, University of Washington, Box 351580, U.W., Seattle, WA 98195-1580, USA}

\author[0000-0002-4338-6586]{P. Hoeflich}
\affil{Department of Physics, Florida State University, 77 Chieftan Way, Tallahassee, FL 32306, USA}

\author[0000-0002-1966-3942]{R. P. Kirshner}
\affil{Harvard-Smithsonian Center for Astrophysics, 60 Garden Street, Cambridge, MA 02138, USA}
\affil{Gordon and Betty Moore Foundation, 1661 Page Mill Road, Palo Alto, CA 94304, USA}

\author[0000-0002-6650-694X]{K. Krisciunas}
\affil{George P. and Cynthia Woods Mitchell Institute for Fundamental Physics and Astronomy, Department of Physics and Astronomy, Texas A\&M University, College Station, TX, 77843, USA}

\author{G. H. Marion}
\affiliation{Department of Astronomy, The University of Texas at Austin, 1 University Station C1400, Austin, TX 78712-0259, USA}

\author[0000-0003-2535-3091]{N. Morrell}
\affil{Carnegie Observatories, Las Campanas Observatory, Casilla 601, La Serena, Chile}

\author[0000-0003-0554-7083]{S. E. Persson}
\affil{Observatories of the Carnegie Institution for Science, 813 Santa Barbara St, Pasadena, CA 91101, USA}

\author[0000-0003-2734-0796]{M. M. Phillips}
\affil{Carnegie Observatories, Las Campanas Observatory, Casilla 601, La Serena, Chile}

\author[0000-0002-5571-1833]{M. D. Stritzinger}
\affil{Department of Physics and Astronomy, Aarhus University, Ny Munkegade 120, DK-8000 Aarhus C, Denmark}

\author[0000-0002-8102-181X]{N. B. Suntzeff}
\affil{George P. and Cynthia Woods Mitchell Institute for Fundamental Physics and Astronomy, Department of Physics and Astronomy, Texas A\&M University, College Station, TX, 77843, USA}

\author[0000-0002-2387-6801]{F. Taddia}
\affil{Department of Physics and Astronomy, Aarhus University, Ny Munkegade 120, DK-8000 Aarhus C, Denmark}

\begin{abstract}

Unburned carbon is potentially a powerful probe of Type Ia supernova (SN) explosion mechanisms.  We present comprehensive optical and near-infrared (NIR) data on the ``transitional" Type Ia SN~2015bp. An early NIR spectrum ($t = -$9.9 days with respect to $B$-band maximum) displays a striking \ion{C}{1} $\lambda1.0693\,\mu \rm{m}$ line at $11.9 \times 10^3$~km s$^{-1}$, distinct from the prominent \ion{Mg}{2} $\lambda1.0927\,\mu \rm{m}$ feature, which weakens toward maximum light. SN~2015bp also displays a clear \ion{C}{2} $\lambda$6580\AA~notch early ($t = -10.9$ days) at $13.2 \times 10^3$~km s$^{-1}$, consistent with our NIR carbon detection. At $M_B = -$18.46, SN 2015bp is less luminous than a normal SN Ia and, along with iPTF13ebh, is the second member of the transitional subclass to display prominent early-time NIR carbon absorption. We find it unlikely that the \ion{C}{1} feature is misidentified \ion{He}{1} $\lambda1.0830\,\mu\rm{m}$ because this feature grows weaker toward maximum light, while the helium line produced in some double-detonation models grows stronger at these times. Intrigued by these strong NIR carbon detections, but lacking NIR data for other SNe Ia, we investigated the incidence of optical carbon in the sample of nine transitional SNe Ia with early-time data ($t \lesssim-$4 days). We find that four display \ion{C}{2} $\lambda$6580\,\AA, while two others show tentative detections, in line with the SN Ia population as a whole. We conclude that at least $\sim$50\% of transitional SNe Ia in our sample do not come from sub-Chandrasekhar mass explosions due to the clear presence of carbon in their NIR and optical spectra.

%ORIGINAL: The common occurrence of carbon in the spectra of the faint transitional SN Ia subclass argues against a sub-Chandrasekhar-mass explosion as their origin, despite recent theoretical work suggesting this must be the case.

\end{abstract}

 
\section{Introduction}\label{sec:intro}

Type Ia supernovae (SNe Ia) are important ``standardizable'' candles, following a width-luminosity relationship where fainter events have faster declining light curves \citep{Phillips93}.  Using the resulting empirically calibrated luminosities, SNe Ia have provided a direct measure of the expansion history of the Universe, ultimately leading to the discovery of the accelerating expansion and ``dark energy'' \citep{reiss1998,perlmutter1999}.  However, supernova cosmology is currently limited by systematic errors \citep[e.g.,][]{Conley11,Suzuki12, brout2019}, and further progress may mean shifting observations to the near-infrared (NIR) where SNe Ia are even better standard candles than in the optical \citep[e.g.,][]{Krisciunas04,Folatelli10,Kattner12, avelino2019}.  A deeper understanding of the physics of SNe Ia is also vital both for mitigating systematic errors in future cosmology experiments and for the ultimate quest to understand the late stages of stellar evolution.

One potential, critical probe of SN Ia explosions is the incidence, quantity, and distribution of leftover carbon.  There is a general consensus that SNe Ia result from the thermonuclear explosion of a carbon-oxygen white dwarf \citep{Hoyle60}.  Given this, carbon is the only direct probe of primordial material from the progenitor system, as oxygen is produced in carbon burning during the explosion. Specific explosion parameters and the degree of mixing are key ingredients in determining whether leftover carbon is expected for a given explosion model.   For instance, the pure deflagration W7 model of \citet{Nomoto84} leaves substantial carbon behind, while delayed-detonation models have nearly complete carbon burning for normal SNe Ia \citep{kasen2009}, with increasing amounts of unburned carbon for fainter events \citep{hoflich2002}.  Both the pulsating class of delayed-detonation models \citep{Hoflich95} and violent merger models \citep{Pakmor12} also have a large mass fraction of carbon.  The class of sub-Chandrasekhar double-detonation models, however, are characterized by the nearly complete lack of surviving carbon in recent work \citep[e.g.,][]{Shen18,polin2019}, although a small amount may remain below the outer layer of iron-group elements during a surface helium detonation \citep{Fink10}.  Connecting observed carbon to model expectations is of course nontrivial, but clear observational samples play an important role in distinguishing between viable SN Ia explosion mechanisms.

The identification of carbon in SNe Ia began with individual detections in normal \citep[e.g.,][]{Branch03,Garavini04,Thomas07}, faint \citep[e.g.,][]{hoflich2002,Taubenberger08} and super-Chandrasekhar \citep[e.g.,][]{Howell06,Scalzo10,Silverman11} events, followed by several large-sample studies focused on the optical and the \ion{C}{2} $\lambda6580\,\mbox{\AA}$ feature, which appears on the red shoulder of the prominent \ion{Si}{2} $\lambda6355\,\mbox{\AA}$ absorption line \citep{Thomas11,parrent2011,folatelli2012,silverman2012,maguire2014}.  In these studies, $\sim20$--40\% of pre-maximum spectra showed \ion{C}{2} features, with the fraction increasing the earlier a spectrum is taken before maximum light.  In addition to this observational detection bias, it is also likely that high-velocity 
\ion{C}{2} $\lambda6580\,\mbox{\AA}$ is undetectable owing to the proximity and strength of the \ion{Si}{2} line \citep[see, e.g., ][]{Thomas11,folatelli2012}.
There is also evidence that SNe Ia with bluer optical and ultraviolet (UV) colors are more likely to exhibit carbon features \citep{Thomas11,folatelli2012,silverman2012,milnebrown2013}, which may suggest that carbon is preferentially seen in low-metallicity progenitor systems \citep{Heringer17}.

While most carbon studies have focused on the \ion{C}{2} $\lambda6580\,\mbox{\AA}$ line, there is observational evidence that  NIR \ion{C}{1} $\lambda1.0693$\,$\mu$m may be more prevalent, although the datasets are still sparse.  Using the remarkable early-time and high-cadence dataset for SN~2011fe, \citet{hsiao2013} noted the strength of the NIR \ion{C}{1} $\lambda1.0693$\,$\mu$m line increased toward maximum light, and suggested this may be a consequence of the recombination of \ion{C}{2} to \ion{C}{1} as the ejecta cool.  Since then, several other studies have noted a similar trend for individual objects \citep{iptf13ebhhsiao,marion2015,CSPII2}, although it should be noted that spectrum synthesis codes such as \texttt{SYNAPPS} \citep{Thomas11} are necessary to identify the \ion{C}{1} $\lambda1.0693$\,$\mu$m line in the ``shoulder" of the stronger \ion{Mg}{2} $\lambda1.0927\,\mu$m line for normal SNe Ia.  

Perhaps the most intriguing case of NIR \ion{C}{1} $\lambda1.0693$\,$\mu$m is that of iPTF13ebh \citep{iptf13ebhhsiao}, which showed a very prominent, conspicuous absorption feature blueward of \ion{Mg}{2} $\lambda1.0927\,\mu$m in early-time data ($t \lesssim -$10 d) which grew weaker toward maximum light, in contrast to the subtler appearance of \ion{C}{1} in normal SNe Ia.   iPTF13ebh is a ``transitional" SN Ia, a faint \citep[and rare;][]{Ashall16} subclass intermediate between normal and the even more subluminous SN 1991bg-like subclass.  To go with their subluminous nature, transitional SNe Ia have fast decline rates (typically $\Delta m_{15}(B) \approx 1.5$--2.0 mag), but lack the strong \ion{Ti}{2} feature seen in SN 1991bg-like events \citep[$\sim4000$--4500\,\AA;][]{filippenko1992}.  They also exhibit a NIR primary maximum which occurs before the $B$-band maximum epoch and exhibit a NIR secondary maximum, again in contrast to the more underluminous SN 1991bg-like SNe Ia \citep{krisciunas09}.
Recent modeling efforts have suggested that faint-and-fast declining SNe Ia, such as the transitional subclass, are largely produced by sub-Chandrasekhar explosions \citep{Blondin17,Goldstein18} -- in tension with the clear detection of \ion{C}{1} in iPTF13ebh, as carbon should be nearly nonexistent in sub-Chandrasekhar explosions.  Meanwhile, abundance stratification studies of other transitional SNe Ia (SN~1986G, SN~2007on, and SN~2011iv) have been consistent with Chandrasekhar mass delayed-detonation models \citep{Ashall16,Ashall18}.  Further observational and theoretical work on transitional SNe Ia will help elucidate their progenitors and explosions.

Here we report on a second transitional SN Ia with strong early-time NIR carbon, SN~2015bp \citep[originally studied by][]{Srivastav17}.  We also undertake an expanded search for optical carbon in the present sample of 9 transitional SNe Ia with early-time data.
In Section~\ref{sec:obsred} we present our SN~2015bp observations, while in Section~\ref{sec:phtprop} and Section~\ref{sec:specprop} we describe its photometric and spectroscopic properties, respectively.  Section~\ref{sec:carbcomp} discusses both the NIR and optical detection of carbon in SN~2015bp and compares it with other prominent examples in the literature.  Intrigued by the strength of NIR carbon in both iPTF13ebh and SN~2015bp, we then (Sec.~\ref{sec:cii_inc}) search for early optical carbon in the full sample of transitional SNe Ia in the literature.  Given the incidence of both NIR and optical carbon in the transitional SN Ia subclass, we discuss the implications for various explosion models in Section~\ref{sec:disc}, before concluding in Section~\ref{sec:conc}.  

\section{Observations}\label{sec:obsred}

\subsection{Photometric Observations}\label{sec:phtobs}
\begin{deluxetable*}{ccccccccc}
\tablecaption{Optical photometry of SN~2015bp in the natural Swope system. The phase is with respect to the time of $B$-band maximum. \label{tab:photdatatab}}
\tablehead{\colhead{Date} & \colhead{MJD} & \colhead{Phase (d)} & \colhead{$u$ (mag)} & \colhead{$B$ (mag)} & \colhead{$g$ (mag)} & \colhead{$V$ (mag)} & \colhead{$r$ (mag)} & \colhead{$i$ (mag)}}
\startdata
2015-03-19.34 & 57100.34 & $-$11.38 & $16.560 \pm 0.012$ & $16.339 \pm 0.012$ & $17.435 \pm 0.019$ & $16.403 \pm 0.008$ & $16.372 \pm 0.011$ & $16.547 \pm 0.015$ \\
2015-03-20.35 & 57101.35 & $-$10.37 & $15.977 \pm 0.005$ & $15.854 \pm 0.006$ & $16.645 \pm 0.010$ & $15.872 \pm 0.005$ & $15.862 \pm 0.005$ & $16.022 \pm 0.007$ \\
2015-03-21.32 & 57102.32 & $-$9.40 & $15.541 \pm 0.006$ & $15.455 \pm 0.006$ & $16.019 \pm 0.007$ & $15.447 \pm 0.005$ & $15.437 \pm 0.005$ & $15.581 \pm 0.006$ \\
2015-03-22.36 & 57103.36 & $-$8.36 & ... & $15.082 \pm 0.013$ & $15.509 \pm 0.013$ & $15.093 \pm 0.011$ & $15.057 \pm 0.012$ & $15.206 \pm 0.018$ \\
2015-03-28.34 & 57109.34 & $-$2.38 & $13.966 \pm 0.005$ & $13.937 \pm 0.005$ & $14.468 \pm 0.006$ & $13.902 \pm 0.005$ & $13.893 \pm 0.006$ & $14.200 \pm 0.005$ \\
2015-03-29.30 & 57110.30 & $-$1.42 & $13.941 \pm 0.010$ & $13.883 \pm 0.009$ & $14.448 \pm 0.008$ & $13.859 \pm 0.007$ & $13.838 \pm 0.009$ & $14.190 \pm 0.010$ \\
2015-03-30.26 & 57111.26 & $-$0.46 & $13.899 \pm 0.009$ & $13.838 \pm 0.008$ & $14.478 \pm 0.006$ & $13.822 \pm 0.007$ & $13.808 \pm 0.006$ & $14.238 \pm 0.009$ \\
2015-03-31.35 & 57112.35 & 0.63 & $13.886 \pm 0.006$ & $13.805 \pm 0.006$ & $14.555 \pm 0.005$ & $13.799 \pm 0.005$ & $13.800 \pm 0.007$ & $14.274 \pm 0.007$ \\
2015-04-01.31 & 57113.31 & 1.59 & $13.892 \pm 0.011$ & $13.792 \pm 0.010$ & $14.629 \pm 0.007$ & $13.785 \pm 0.007$ & $13.738 \pm 0.008$ & $14.300 \pm 0.008$ \\
2015-04-05.29 & 57117.29 & 5.57 & $14.134 \pm 0.010$ & $13.855 \pm 0.007$ & $15.131 \pm 0.010$ & $13.943 \pm 0.007$ & $13.870 \pm 0.008$ & $14.467 \pm 0.011$ \\
2015-04-09.26 & 57121.26 & 9.54 & $14.582 \pm 0.008$ & $14.130 \pm 0.008$ & $15.734 \pm 0.013$ & $14.294 \pm 0.008$ & $14.199 \pm 0.008$ & $14.811 \pm 0.009$ \\
2015-04-10.27 & 57122.27 & 10.55 & $14.725 \pm 0.006$ & $14.199 \pm 0.007$ & $15.890 \pm 0.009$ & $14.413 \pm 0.008$ & $14.273 \pm 0.009$ & $14.852 \pm 0.010$ \\
2015-04-11.29 & 57123.29 & 11.57 & $14.874 \pm 0.007$ & $14.280 \pm 0.006$ & $16.056 \pm 0.009$ & $14.535 \pm 0.008$ & $14.310 \pm 0.006$ & $14.842 \pm 0.007$ \\
2015-04-13.32 & 57125.32 & 13.60 & $15.166 \pm 0.010$ & $14.455 \pm 0.008$ & $16.413 \pm 0.014$ & $14.771 \pm 0.006$ & $14.372 \pm 0.005$ & $14.798 \pm 0.007$ \\
2015-04-14.29 & 57126.29 & 14.57 & $15.310 \pm 0.007$ & $14.521 \pm 0.006$ & $16.552 \pm 0.015$ & $14.903 \pm 0.006$ & $14.400 \pm 0.006$ & $14.767 \pm 0.011$ \\
2015-04-17.24 & 57129.24 & 17.52 & $15.704 \pm 0.010$ & $14.779 \pm 0.007$ & $16.994 \pm 0.015$ & $15.301 \pm 0.006$ & $14.513 \pm 0.006$ & $14.725 \pm 0.009$ \\
2015-04-18.27 & 57130.27 & 18.55 & $15.833 \pm 0.008$ & $14.863 \pm 0.006$ & $17.106 \pm 0.013$ & $15.414 \pm 0.005$ & $14.554 \pm 0.005$ & $14.713 \pm 0.006$ \\
2015-04-19.31 & 57131.31 & 19.59 & $15.956 \pm 0.010$ & $14.959 \pm 0.006$ & $17.196 \pm 0.014$ & $15.526 \pm 0.007$ & $14.626 \pm 0.008$ & $14.715 \pm 0.010$ \\
2015-04-20.28 & 57132.28 & 20.56 & $16.052 \pm 0.010$ & $15.058 \pm 0.009$ & $17.288 \pm 0.015$ & $15.638 \pm 0.006$ & $14.707 \pm 0.005$ & $14.762 \pm 0.007$ \\
2015-04-22.32 & 57134.32 & 22.6 & $16.242 \pm 0.009$ & $15.247 \pm 0.006$ & $17.431 \pm 0.014$ & $15.831 \pm 0.006$ & $14.908 \pm 0.006$ & $14.916 \pm 0.006$ \\
2015-04-23.33 & 57135.33 & 23.61 & $16.339 \pm 0.009$ & $15.346 \pm 0.009$ & $17.488 \pm 0.016$ & $15.930 \pm 0.007$ & $15.014 \pm 0.006$ & $15.016 \pm 0.007$ \\
2015-04-24.31 & 57136.31 & 24.59 & $16.420 \pm 0.010$ & $15.427 \pm 0.007$ & $17.565 \pm 0.013$ & $15.992 \pm 0.006$ & $15.105 \pm 0.008$ & $15.107 \pm 0.008$ \\
2015-04-25.27 & 57137.27 & 25.55 & $16.466 \pm 0.011$ & $15.502 \pm 0.010$ & $17.629 \pm 0.016$ & $16.051 \pm 0.008$ & $15.191 \pm 0.010$ & $15.195 \pm 0.010$ \\
2015-04-26.26 & 57138.26 & 26.54 & $16.524 \pm 0.008$ & $15.556 \pm 0.007$ & $17.688 \pm 0.014$ & $16.122 \pm 0.005$ & $15.268 \pm 0.005$ & $15.285 \pm 0.007$ \\
2015-04-27.25 & 57139.25 & 27.53 & $16.570 \pm 0.008$ & $15.625 \pm 0.006$ & $17.735 \pm 0.013$ & $16.175 \pm 0.007$ & $15.333 \pm 0.005$ & $15.358 \pm 0.007$ \\
2015-04-28.24 & 57140.24 & 28.52 & $16.603 \pm 0.008$ & $15.678 \pm 0.006$ & $17.810 \pm 0.013$ & $16.231 \pm 0.006$ & $15.408 \pm 0.006$ & $15.423 \pm 0.008$ \\
2015-04-30.26 & 57142.26 & 30.54 & $16.692 \pm 0.011$ & $15.775 \pm 0.009$ & $17.870 \pm 0.019$ & $16.321 \pm 0.009$ & $15.518 \pm 0.007$ & $15.562 \pm 0.009$ \\
2015-05-02.24 & 57144.24 & 32.52 & $16.795 \pm 0.013$ & $15.869 \pm 0.007$ & $17.914 \pm 0.027$ & $16.383 \pm 0.007$ & $15.624 \pm 0.007$ & $15.672 \pm 0.006$ \\
2015-05-06.30 & 57148.30 & 36.58 & $16.889 \pm 0.014$ & $16.000 \pm 0.010$ & $18.027 \pm 0.034$ & $16.509 \pm 0.009$ & $15.812 \pm 0.008$ & $15.888 \pm 0.008$ \\
2015-05-09.26 & 57151.26 & 39.54 & $16.960 \pm 0.011$ & $16.089 \pm 0.007$ & $18.122 \pm 0.023$ & $16.598 \pm 0.006$ & $15.930 \pm 0.005$ & $16.006 \pm 0.006$ \\
2015-05-11.26 & 57153.26 & 41.54 & $17.024 \pm 0.011$ & $16.177 \pm 0.009$ & $18.166 \pm 0.016$ & $16.655 \pm 0.009$ & $16.018 \pm 0.009$ & $16.125 \pm 0.010$ \\
2015-05-13.25 & 57155.25 & 43.53 & $17.079 \pm 0.011$ & $16.241 \pm 0.012$ & $18.251 \pm 0.017$ & $16.712 \pm 0.008$ & $16.099 \pm 0.010$ & $16.225 \pm 0.013$ \\
2015-05-15.24 & 57157.24 & 45.52 & $17.127 \pm 0.011$ & $16.306 \pm 0.008$ & $18.314 \pm 0.024$ & $16.758 \pm 0.012$ & $16.130 \pm 0.009$ & $16.311 \pm 0.009$ \\
2015-05-17.28 & 57159.28 & 47.56 & $17.153 \pm 0.009$ & $16.358 \pm 0.007$ & $18.340 \pm 0.017$ & $16.816 \pm 0.007$ & $16.246 \pm 0.008$ & $16.378 \pm 0.010$ \\
2015-05-19.25 & 57161.25 & 49.53 & $17.217 \pm 0.009$ & $16.421 \pm 0.009$ & $18.453 \pm 0.021$ & $16.858 \pm 0.009$ & $16.318 \pm 0.010$ & $16.453 \pm 0.013$ \\
2015-05-21.24 & 57163.24 & 51.52 & $17.250 \pm 0.008$ & $16.470 \pm 0.009$ & $18.468 \pm 0.017$ & $16.878 \pm 0.007$ & $16.393 \pm 0.006$ & $16.558 \pm 0.008$ \\
2015-05-24.20 & 57166.20 & 54.48 & $17.309 \pm 0.009$ & $16.554 \pm 0.008$ & $18.543 \pm 0.017$ & $16.967 \pm 0.006$ & $16.516 \pm 0.006$ & $16.677 \pm 0.009$ \\
2015-06-03.20 & 57176.20 & 64.48 & $17.562 \pm 0.027$ & $16.834 \pm 0.015$ & $18.922 \pm 0.071$ & $17.177 \pm 0.014$ & $16.880 \pm 0.011$ & $17.049 \pm 0.014$ \\
2015-06-05.23 & 57178.23 & 66.51 & $17.540 \pm 0.024$ & $16.893 \pm 0.015$ & $18.928 \pm 0.049$ & $17.189 \pm 0.012$ & $16.961 \pm 0.011$ & $17.139 \pm 0.015$ \\
2015-06-06.23 & 57179.23 & 67.51 & $17.530 \pm 0.023$ & $16.909 \pm 0.015$ & $19.077 \pm 0.148$ & $17.218 \pm 0.015$ & $17.013 \pm 0.013$ & $17.176 \pm 0.016$ \\
2015-06-08.19 & 57181.19 & 69.47 & $17.570 \pm 0.010$ & $16.996 \pm 0.009$ & $18.945 \pm 0.029$ & $17.247 \pm 0.008$ & $17.081 \pm 0.009$ & $17.265 \pm 0.011$ \\
2015-06-09.19 & 57182.19 & 70.47 & $17.595 \pm 0.011$ & $16.982 \pm 0.007$ & $19.059 \pm 0.026$ & $17.265 \pm 0.007$ & $17.109 \pm 0.007$ & $17.272 \pm 0.010$ \\
2015-08-15.03 & 57249.03 & 137.31 & $18.791 \pm 0.014$ & $18.616 \pm 0.018$ & $20.877 \pm 0.122$ & $18.569 \pm 0.012$ & $19.250 \pm 0.027$ & $19.072 \pm 0.031$ \\
\enddata
\end{deluxetable*}

\begin{deluxetable*}{cccccc}
\tablecaption{NIR photometry of SN~2015bp. Phase is with respect to $B$-band maximum. \label{tab:yjhdatatab}}
\tablehead{\colhead{Date} & \colhead{MJD} & \colhead{Phase (d)} & \colhead{$Y$ (mag)} & \colhead{$J$ (mag)} & \colhead{$H$ (mag)}}
\startdata
2015-03-28.34 & 57109.34 & $-$2.38 & $14.024 \pm 0.005$ & $13.975 \pm 0.005$ & $14.106 \pm 0.005$ \\
2015-03-30.26 & 57111.26 & $-$0.46 & $14.055 \pm 0.005$ & $13.984 \pm 0.005$ & $14.134 \pm 0.005$ \\
2015-03-31.35 & 57112.35 & 0.63 & ... & $14.040 \pm 0.005$ & $14.201 \pm 0.005$ \\
2015-04-01.31 & 57113.31 & 1.59 & ... & $14.105 \pm 0.008$ & $14.259 \pm 0.005$ \\
2015-04-02.27 & 57114.27 & 2.55 & $14.270 \pm 0.006$ & $14.208 \pm 0.005$ & $14.309 \pm 0.006$ \\
2015-04-03.30 & 57115.30 & 3.58 & $14.350 \pm 0.007$ & $14.300 \pm 0.007$ & $14.356 \pm 0.007$ \\
2015-04-08.26 & 57120.26 & 8.54 & $14.662 \pm 0.005$ & $15.195 \pm 0.005$ & $14.527 \pm 0.005$ \\
2015-04-25.27 & 57137.27 & 25.55 & $14.206 \pm 0.005$ & $15.367 \pm 0.005$ & $14.725 \pm 0.005$ \\
2015-04-29.21 & 57141.21 & 29.49 & $14.488 \pm 0.007$ & $15.832 \pm 0.009$ & $15.050 \pm 0.007$ \\
2015-05-02.24 & 57144.24 & 32.52 & $14.703 \pm 0.005$ & $16.138 \pm 0.007$ & $15.224 \pm 0.007$ \\
2015-05-28.18 & 57170.18 & 58.46 & $16.265 \pm 0.007$ & $18.187 \pm 0.034$ & $16.526 \pm 0.020$ \\
2015-05-30.17 & 57172.17 & 60.45 & $16.383 \pm 0.008$ & $18.325 \pm 0.034$ & $16.637 \pm 0.016$ \\
\enddata
\end{deluxetable*}

SN~2015bp was discovered on (UT dates are used throughout this paper) 2015 March 16, 11:45:36 (JD = 2,457,097.99) at $\alpha$(J2000) = $15^{\rm m}05^{\rm m}30.07^{\rm s}$, $\delta$(J2000) = $+01^\circ38'02.40''$ by the Catalina Real-Time Transient Survey \citep[CRTS;][]{crts} at $V=19.2$\,mag, $39''$ offset from the nearby S0 galaxy NGC~5839 (redshift $z=0.004069$). Spectroscopically, it was initially classified as a pre-maximum SN 1991bg-like SN Ia \citep{Jha15}, and it evolved like a normal SN Ia, except that it displayed a fast-declining light curve.

Following discovery, observations were initiated by the Carnegie Supernova Project~II \citep[CSP-II;][]{CSPII1,CSPII2}, and optical photometric observations began on 2015 March 19 using the Henrietta Swope 1\,m telescope at Las Campanas Observatory in Chile. The Swope observations continued until 2015 August 15, obtaining 45 epochs in the CSP-II optical Johnson (\textit{BV}) and Sloan (\textit{ugri}) filters.  Photometry in the Swope natural system \citep[as described  by][]{krisciunas2017} is logged in Table~\ref{tab:photdatatab} and the respective light curves are plotted in Figure~\ref{fig:lightcurve}. Once SN~2015bp had sufficiently faded in March 2016, host-galaxy reference images were obtained for template subtraction. All images were reduced in the manner described by \cite{CSPII1}, where point-spread-function (PSF) photometry of the SN (on the difference images) was computed with respect to a local sequence of stars that were calibrated to \cite{landolt92} and \cite{smith2002} standard-star fields. 

Retrocam on the du Pont 2.5\,m telescope located at Las Campanas Observatory was also used to obtain 11 epochs of NIR photometric data in the \textit{YJH} filters starting on 2015 March 28 and ending on 2015 May 30. These observations are logged in Table~\ref{tab:yjhdatatab}. The images were reduced as described by \cite{contreras2010}. The \textit{JH} bands were both calibrated using the \cite{persson1998} system of standard stars. The \textit{Y} band was calibrated using a set of Persson standards as described by \cite{krisciunas2017}.

\subsection{Spectroscopic Observations}\label{sec:spcobs}

\begin{deluxetable*}{cccccc}
\tablecaption{Journal of optical spectroscopic observations. \label{tab:opspectab}}
\tablehead{\colhead{UT Date} & \colhead{MJD} & \colhead{Instrument} & \colhead{$t_{\rm{max}}(B)^a$} & \colhead{$t_{\rm{int}}^b$} \\ \colhead{} & \colhead{} & \colhead{} & \colhead{$(\rm{days})$} & \colhead{$(\rm{min})$} }
\startdata
 2015-03-20.32 & 57101.3 & EFOSC & $-$10.9 & 45.0 \\
 2015-03-27.16 & 57108.2 & ALFOSC & $-$4.1 & 25.0 \\
 2015-03-28.38 & 57109.4 & EFOSC & $-$2.8 & 10.0 \\
 2015-03-31.00 & 57112.0 & HFOSC & $-$0.2 & $^c$ \\
 2015-04-02.00 & 57114.0 & HFOSC & 1.8 & $^c$ \\
 2015-04-03.00 & 57115.0 & HFOSC & 2.8 & $^c$ \\
 2015-04-05.00 & 57117.0 & HFOSC & 4.8 & $^c$ \\
 2015-04-10.34 & 57122.3 & EFOSC & 10.1 & 10.0 \\
 2015-04-13.11 & 57125.1 & ALFOSC & 12.9 & 30.0 \\
 2015-04-19.22 & 57131.2 & EFOSC & 19.0 & 15.0 \\
 2015-04-24.32 & 57136.3 & WFCCD & 24.1 & 10.0 \\
 2015-05-12.03 & 57154.0 & ALFOSC & 41.8 & 30.0 \\
 2015-05-20.40 & 57162.4 & DEIMOS & 50.2 & 15.0 \\
 2015-07-19.03 & 57222.0 & WFCCD & 109.8 & 16.7 \\
 2015-08-14.90 & 57248.9 & ALFOSC & 136.7 & 40.0 \\
 2015-09-15.98 & 57281.0 & EFOSC & 168.8 & 45.0 
\enddata
\tablecomments{
(a)~$t_{\rm max}(B)$ is the time relative to $B_{\rm max}$, in days.
(b)~$t_{\rm int}$ is the total integration time in minutes.
(c)~Data from the HFOSC instrument were published in \cite{Srivastav17} and were acquired from \url{https://sne.space}. They did not include exposure times in their publication.}
\end{deluxetable*}

\begin{deluxetable*}{cccccc}
\tablecaption{Journal of NIR spectroscopic observations. \label{tab:nirspectab}} 
\tablehead{\colhead{UT Date} & \colhead{MJD} & \colhead{Instrument} & \colhead{$t_{\rm{max}}(B)^a$} & \colhead{$t_{\rm{int}}^b$} \\ \colhead{} & \colhead{} & \colhead{} & \colhead{$(\rm{days})$} & \colhead{$(\rm{min})$} }
\startdata
2015-03-21.27 & 57102.3 & SOFI & $-$9.9 & 36.0 \\
2015-03-23.40 & 57104.4 & F2 & $-$7.8 & 17.5 \\
2015-03-29.31 & 57110.3 & SOFI & $-$1.9 & 36.0 \\
2015-04-02.36 & 57114.4 & FIRE & 2.1 & 8.4 \\
2015-04-07.29 & 57119.3 & FIRE & 7.1 & 8.4 \\
2015-04-12.22 & 57124.2 & FIRE & 12.0 & 8.4 \\
2015-04-16.41 & 57128.4 & GNIRS & 16.2 & 20.0 \\
2015-04-19.38 & 57131.4 & IRTF & 19.2 & 47.4 \\
2015-04-23.23 & 57135.2 & GNIRS & 23.0 & 24.0 \\
2015-05-07.36 & 57149.4 & GNIRS & 37.2 & 35.0 \\
2015-05-17.37 & 57159.4 & IRTF & 47.2 & 49.9 \\
2015-06-01.21 & 57174.2 & FIRE & 62.0 & 21.1 \\
\enddata
\tablecomments{(a)~$t_{\rm max}(B)$ is the time relative to $B_{\rm max}$ in days. (b)~$t_{\rm int}$ is the total integration time in minutes.}
\end{deluxetable*}

Optical spectroscopic observations began on 2013 March 20 and were gathered with the low-resolution EFOSC2 spectrograph on the 3.6\,m New Technology Telescope (NTT), the Wide-Field Reimaging CCD Camera (WFCCD) on the du Pont
100\,inch telescope, the Deep Imaging Multi-Object
Spectrograph (DEIMOS; \citealt{deimos}) on the Keck-II 10\,m telescope,  and the Andalucia Faint Object Spectrograph and Camera (ALFOSC) on the Nordic Optical Telescope (NOT). The optical spectroscopic observations are logged in Table~\ref{tab:opspectab} and displayed in Figure~\ref{fig:optspecevol}. We also include some already published spectra from \cite{Srivastav17} which help fill in the post-maximum evolution. The EFOSC2 data were taken as part of the Public ESO Spectroscopic Survey of Transient Objects (PESSTO) project \citep{smartt2015}, and the reduced spectra were downloaded directly from the ESO archive.

NIR spectroscopic observations began on 2015 March 21 using the Son Of ISAAC (SOFI) instrument on the NTT \citep{SOFI}, the Folded-port InfraRed Echellette \citep[FIRE; ][]{simcoe2013} spectrograph on the Magellan Baade telescope, the Gemini Near Infrared Spectrograph \citep[GNIRS;][]{elias1998} on the Gemini North Telescope, SpeX \citep{rayner2003} on the NASA Infrared Telescope Facility (IRTF), and the Flamingos2 spectrograph  \citep[F2;][]{eikenberry2006} on the Gemini South Telescope. These observations are logged in Table~\ref{tab:nirspectab} and displayed in Figure ~\ref{fig:nirspecevol}. The \textrm{SOFI} data were taken during the PESSTO survey \citep{smartt2015}, and we downloaded the reduced spectra directly from the ESO archive. The FIRE, GNIRS, and F2 spectroscopic data were reduced in the manner described by \citet{CSPII2}. 

Once all the spectra were reduced in their respective manners, they were flux scaled to match the photometry. We also accounted for the Milky Way extinction given a $E(B-V)_{\rm{MW}} = 0.0465 \pm 0.0004$\,mag color excess \citep{Schlafly11} with $R_V=3.1$, and the redshift of the host galaxy ($z=0.004069 \pm 0.000017$) by shifting the wavelength of the spectra to their rest frame. We do not correct for any further host-galaxy extinction \citep[consistent with][]{Srivastav17}. 

\section{Photometric Properties}\label{sec:phtprop}

\begin{figure*}[htp]
\begin{center}
\includegraphics[width=1.0\textwidth, trim=1.0cm {0.25cm} {0.0cm} {0.5cm}]{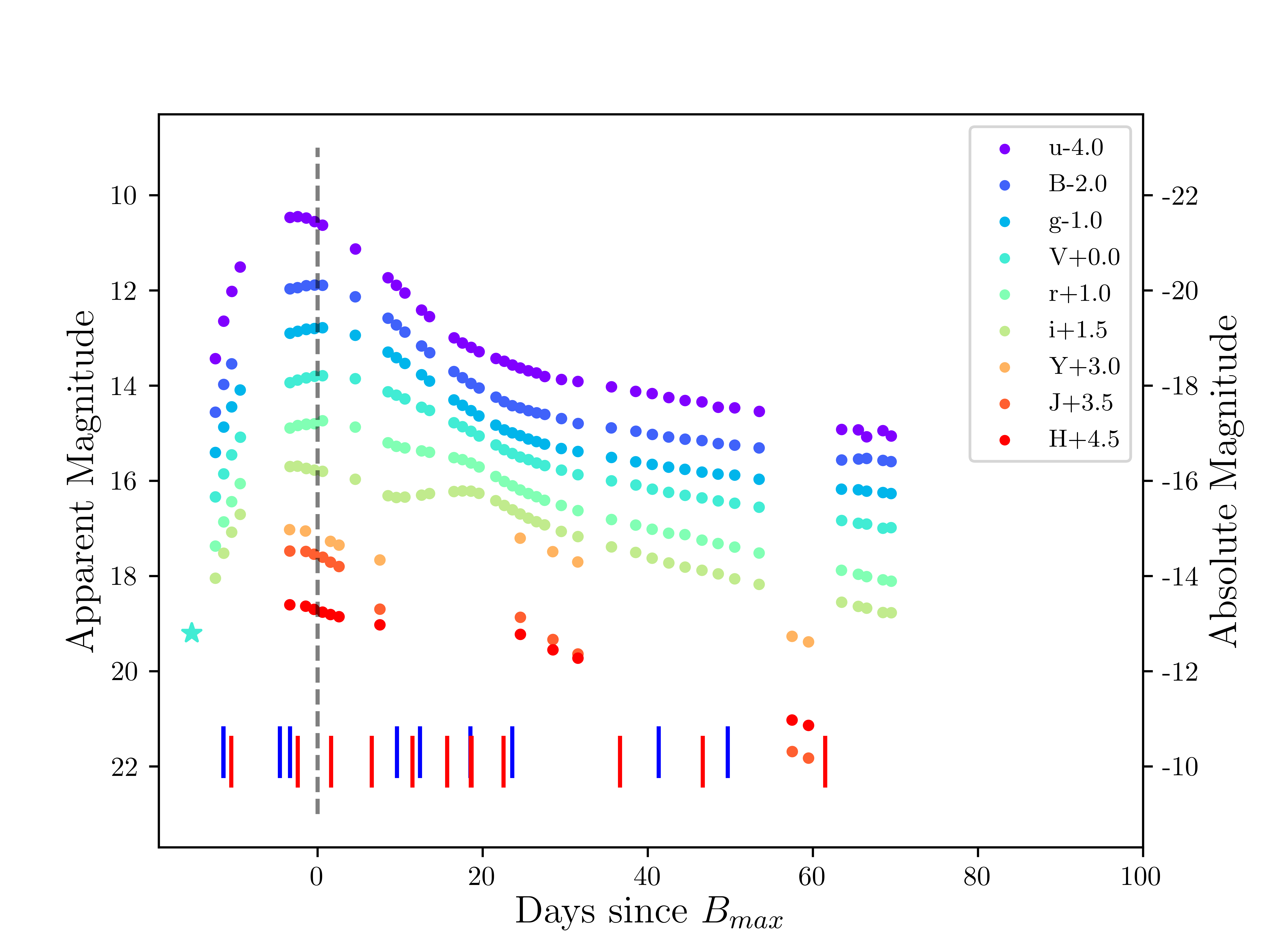}
\caption{Optical and NIR light curves of SN~2015bp.  The light-blue star represents the discovery magnitude from CRTS in the $V$ band. The dashed line represents the time of maximum light in the $B$ band (Table~\ref{tab:sn2015prop}). The blue and red vertical lines represent epochs where optical and NIR spectra were obtained, respectively.}
\label{fig:lightcurve}
\end{center}
\end{figure*}

\begin{figure*}[htp]
\begin{center}
\includegraphics[width=1.0\textwidth, trim=1.0cm {0.25cm} {0.25cm} {-0.0cm}]{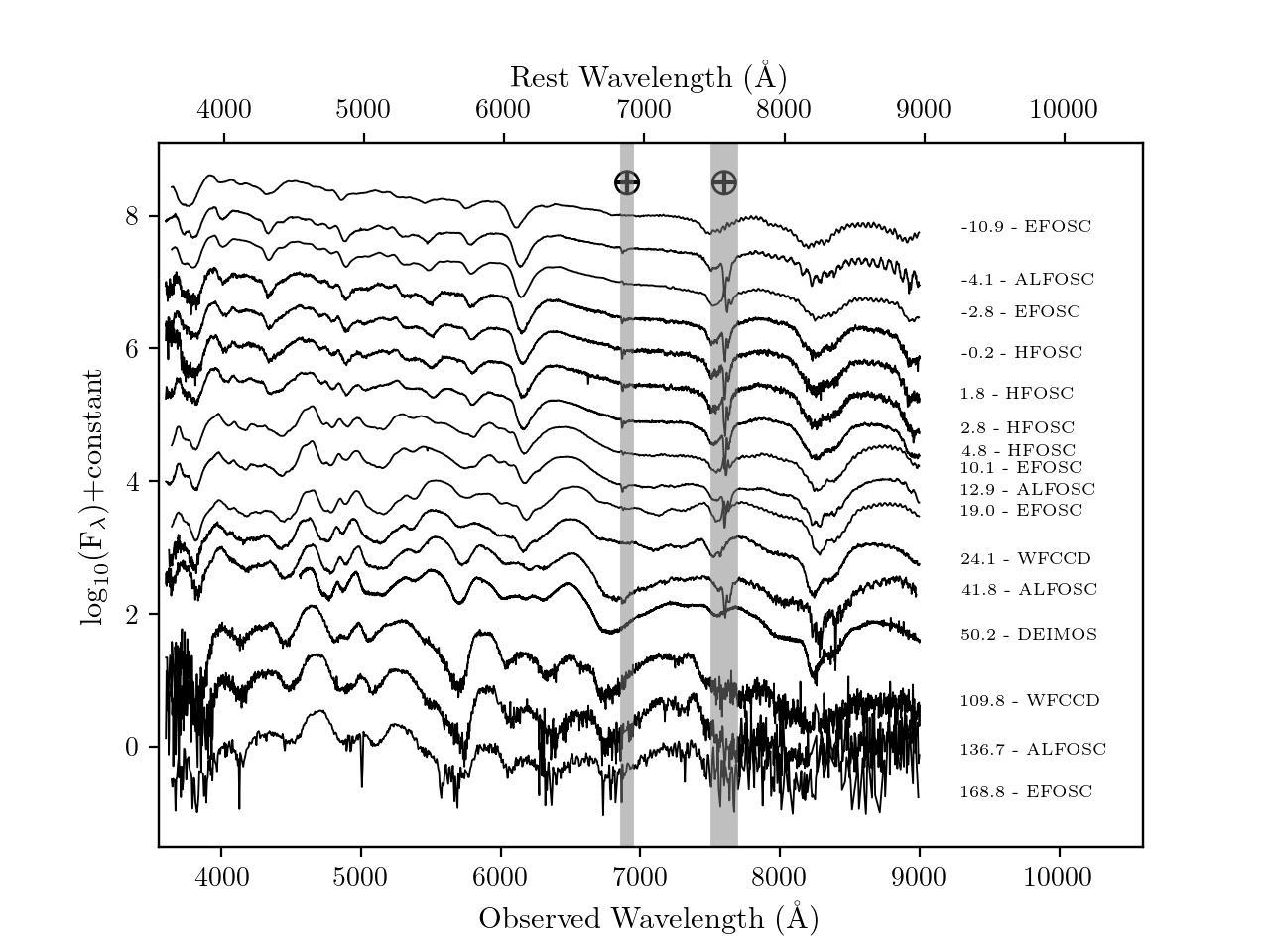}
\caption{Optical spectral evolution of SN~2015bp. Labels are displayed in days relative to $B_{\rm{max}}$.}
\label{fig:optspecevol}
\end{center}
\end{figure*}

\begin{figure*}[htp]
\begin{center}
\includegraphics[width=1.0\textwidth]{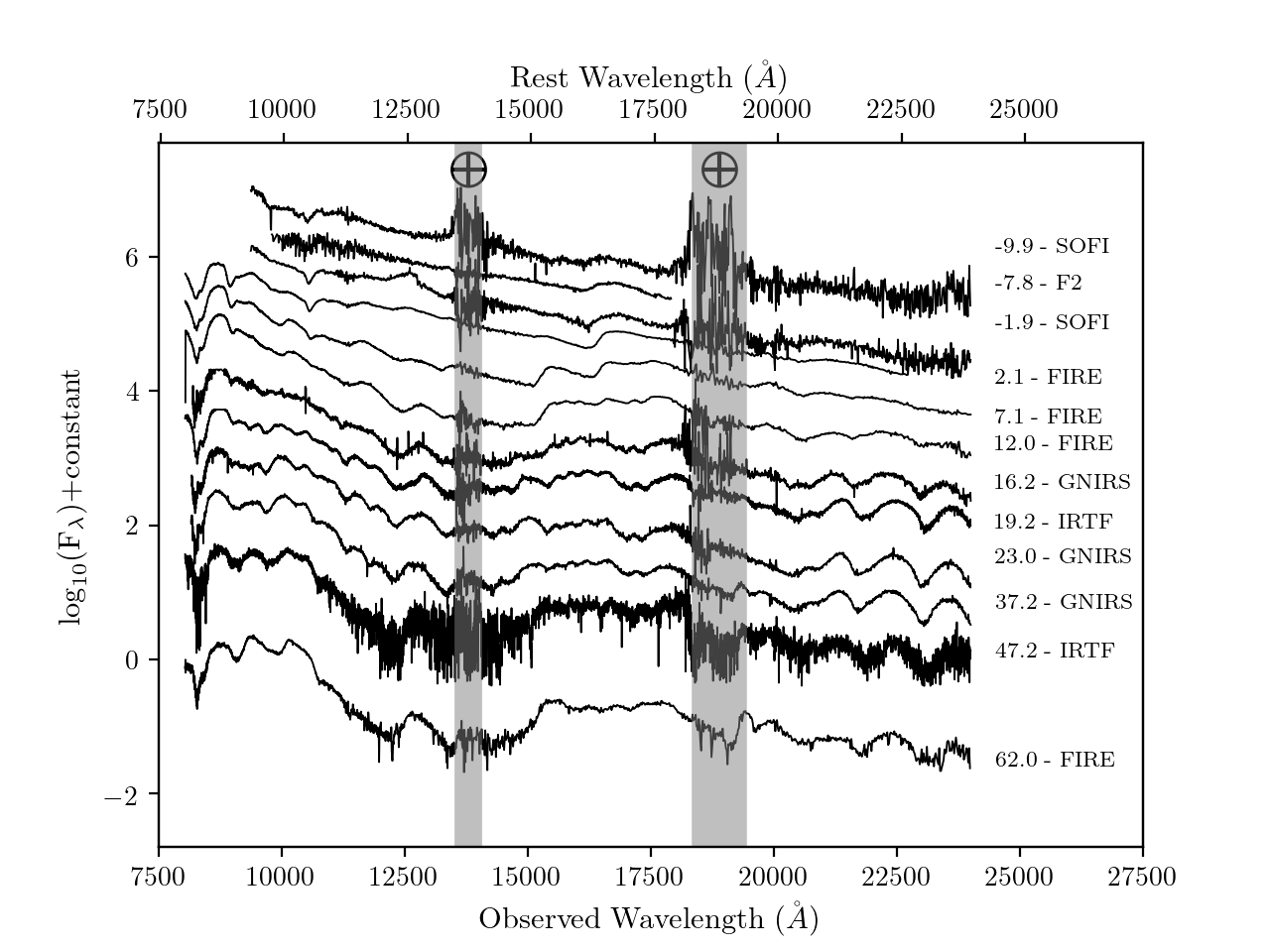}
\caption{NIR spectral evolution of SN~2015bp. Labels are displayed relative to $B_{\rm{max}}$.
}
\label{fig:nirspecevol}
\end{center}
\end{figure*}

\begin{figure}[htp]
\begin{center}
\includegraphics[width=0.5\textwidth, trim=1.0cm {0.25cm} {0.0cm} {0.5cm}]{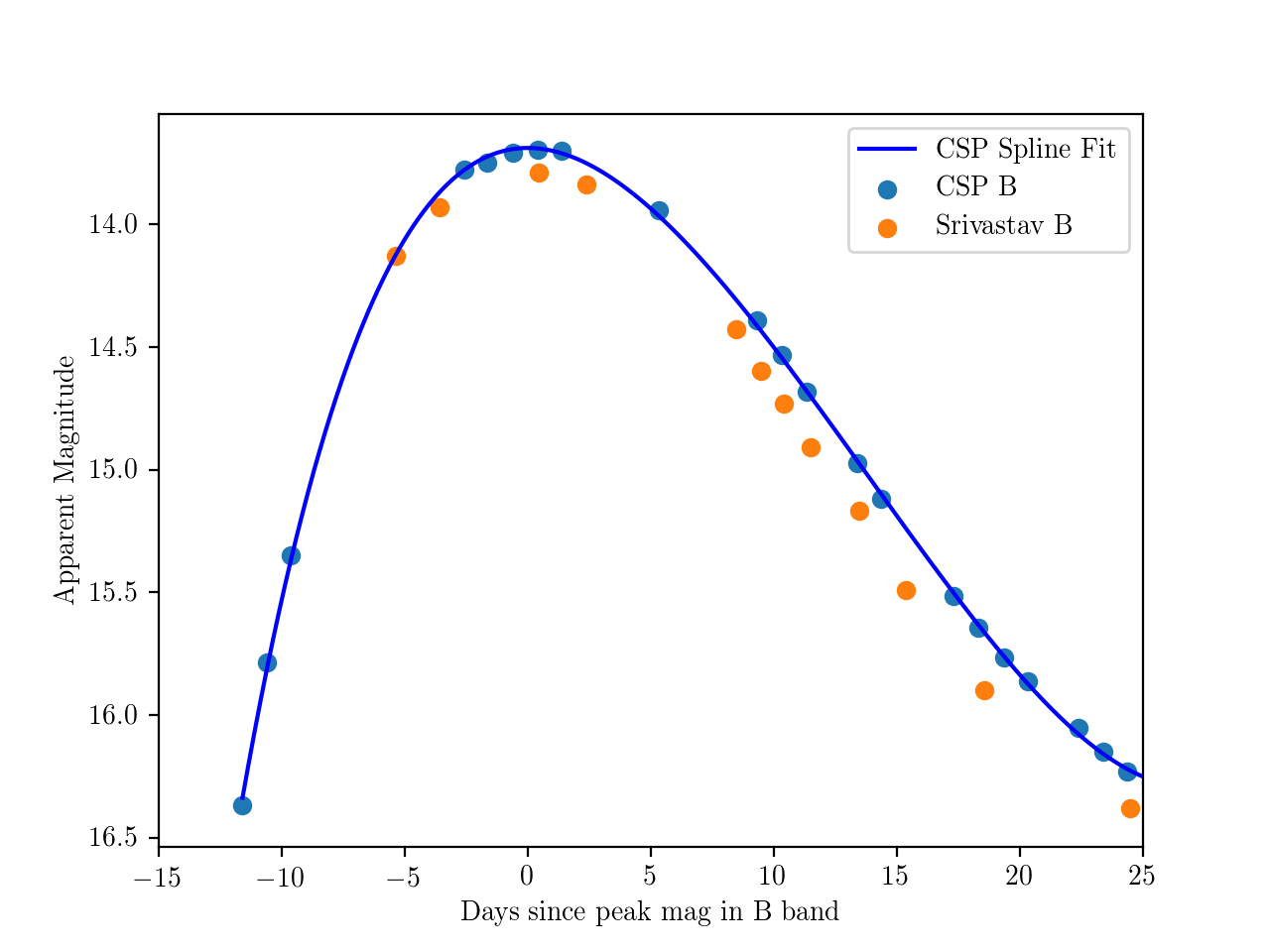}
\caption{Comparison of the $B$-band photometry from \cite{Srivastav17} and the CSP-II $B$ photometry for SN~2015bp, displaying the discrepancy between the $\Delta m _{15}(B)$.  A spline fit to the CSP-II  $B$ data is plotted to guide the eye.} 
\label{fig:srivastavbcomp}
\end{center}
\end{figure}

Optical light-curve data ($BVugri$) were obtained for SN~2015bp between
$t = -11$ days with respect to $B_{\rm{max}}$ to roughly 137 days afterward, whereas the NIR data ($YJH$) span $t = -$3 to 60 days. These data are plotted in Figure \ref{fig:lightcurve} up to day $\sim70$. 
All of the data are logged in Tables \ref{tab:photdatatab} and \ref{tab:yjhdatatab}. 

We fit the light curve using the \texttt{SNooPy} software package \citep[][with $H_0=72$\,km\,s$^{-1}$\,Mpc$^{-1}$ to calibrate the Phillips relation]{Burns11}, and we tabulate the results in Table~\ref{tab:sn2015prop}.  SN~2015bp reached $B_{\rm{max}}$ on JD 2,457,112.72  with an apparent magnitude of $m_{B}=13.69$ after applying reddening corrections due to foreground Milky Way extinction. \texttt{SNooPy} infers a distance modulus of $\mu=32.426 \pm 0.088$\,mag and an absolute magnitude of $M_{B} = -18.73$\,mag.

There is an existing Tully-Fisher based distance measurement to NGC~5839 corresponding to $\mu=32.1 \pm 0.40$\,mag \citep{theureau2007}; which if applied would yield a $M_{B}=-18.46\pm0.41$\,mag.  For this paper, we use the Tully-Fisher-based distance modulus and absolute magnitude going forward, and note that if we adopt the \texttt{SNoopy} distance we would obtain a similar absolute magnitude to within the uncertainties.

Further, we measure a decline rate of $\Delta m_{15}(B) = 1.56 \pm 0.03$\,mag and color-stretch parameter $s_{BV} = 0.671 \pm 0.030$. The color-stretch parameter in particular is a better way of characterizing the light curve for fast-declining SNe Ia \citep[see][]{burns2014}, as the linear decline in the $B$-band occurs earlier than +15\,d for fast-declining SNe, and the $M_B$ versus $\Delta m_{15}(B)$ relation bifurcates in this region.  Using the $s_{BV}$ parameter leads to the fastest-declining events appearing as a continuous tail end of the normal SN Ia population \citep[see e.g.,][]{burns2018,gall2018}.

We recognize that our measured value of $\Delta m_{15}(B)$ is smaller than what \cite{Srivastav17} report in their study of SN~2015bp. In Figure~\ref{fig:srivastavbcomp} we show a comparison of the CSP-II $B$-band light curve (around the time of $B_{\rm{max}}$) compared to the photometry reported by \cite{Srivastav17}. Assuming the $B$-band light curve of \cite{Srivastav17} is on the standard filter system, the S-correction to the Swope natural system is on the $1\%$ level, and does not account for the discrepancy observed.
They report a $\Delta m_{15}(B)$ value at 1.72\,mag, whereas ours is at 1.56\,mag. We performed a direct low-order polynomial fit to the \citet{Srivastav17} $B$-band data which yielded $\Delta m_{15}(B)=1.68\pm 0.03$\,mag.
Despite the discrepancy, both values are consistent with light-curve decline rates for transitional SNe Ia. The data in the current work better sampled the light curve around maximum light and at earlier phases, both of which make us confident in our measurements.  

Transitional SNe Ia have the characteristic in which they reach primary NIR maxima slightly prior to their $t_{\rm{max B}}$ \citep[as discussed in ][]{iptf13ebhhsiao}. Examining the NIR light curves for SN~2015bp, there is evidence that it does exhibit the `transitional' characteristic of having its primary NIR peak before $t_{\rm{max B}}$ along with a  secondary peak.  This is evident primarily in the \textit{i} and \textit{Y} bands. 

\begin{table}
  \begin{threeparttable}
    \caption{Basic properties of SN~2015bp \label{tab:sn2015prop}}
     \begin{tabular}{cc}
     	\hline
        $\alpha$(J2000) & $15^{\rm h}05^{\rm m}30.07^{\rm s}$ \\
        $\delta$(J2000) & $+01^\circ38'02.40''$ \\
        JD$_{\rm{explosion}}^{a}$ & $2,457,093.64\pm1.68$ \\
        JD$_{\rm{discovery}}$ & $2,457,097.99$ \\
        JD$_{\rm{max}}(B)$ & $2,457,112.71$ \\
        $m_{B,\rm{max}}$ & $13.69 \pm 0.01$ mag\\
        $B_{\rm{abs}} (\rm{max})^{b,c}$ & $-18.46 \pm 0.41$ mag\\
        $\Delta m_{15}(B)$ & $1.56 \pm 0.03$ mag\\
        $s_{BV}$ & $0.671 \pm 0.030$ \\
        Host & NGC 5839 \\
        Heliocentric Redshift$^{d}$ & $0.004069 \pm 0.000017$\\
        Distance Modulus$^{e}$ & $32.15\pm0.40$\,mag \\
        Distance Modulus$^{f}$ & $32.426\pm0.088$\,mag \\
        $E(B-V)_{\rm{MW}}$ & $0.0465 \pm 0.0004$\,mag\\
        \hline
    \end{tabular}
    \begin{tablenotes}
      \small
      \item $^{a}$ Derived from the fit of the $v \approx t^{-0.22}$ power law of \cite{pironakar2013} to the \ion{Si}{2} $\lambda 6355\, \mbox{\AA}$ velocity time evolution.
      \item $^{b}$ Peak magnitudes include a reddening correction for foreground Milky Way extinction. 
      \item $^{c}$ Absolute magnitude calculated using the distance modulus from the mean Tully-Fisher relation.
	  \item $^{d}$ \cite{cappellari2011}.
      \item $^{e}$ Distance modulus estimated using the mean Tully-Fisher relation from \cite{theureau2007}.
      \item $^{f}$ Distance modulus estimated using SNooPy; \cite{burns2014}.
    \end{tablenotes}
  \end{threeparttable}
\end{table}

\section{Spectroscopic Properties}\label{sec:specprop}

\begin{figure*}[htp]
\begin{center}
\includegraphics[width=1.0\textwidth]{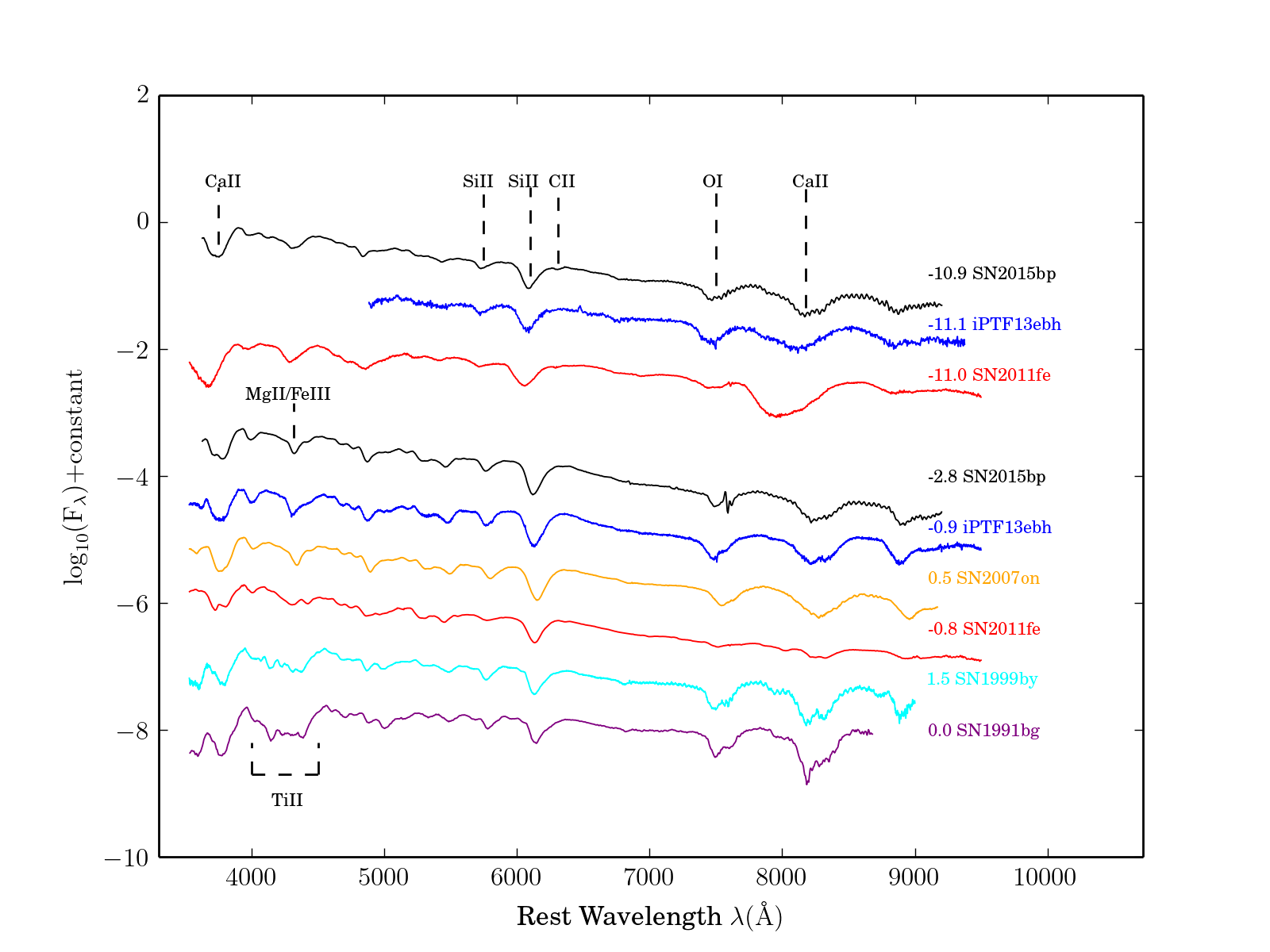}
\caption{Comparison of optical spectra of SN~2015bp with  those of other notable SNe Ia, both at a pre-maximum and near-maximum light phase. Absorption features discussed in the text are noted.
}
\label{fig:spectracomp}
\end{center}
\end{figure*}

\begin{figure}[htp]
\begin{center}
\includegraphics[width=0.5\textwidth]{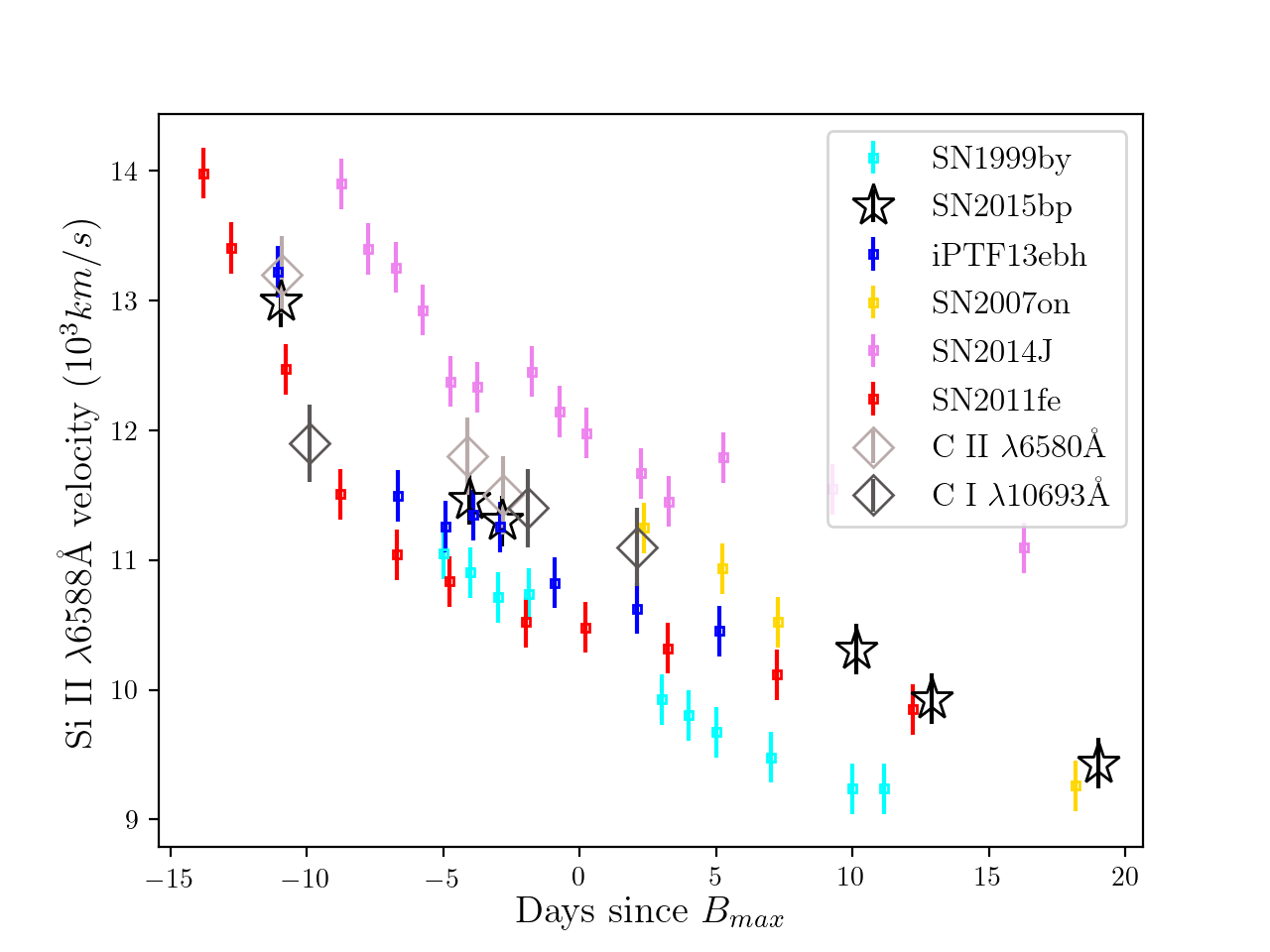}
\caption{\ion{Si}{2} $\lambda 6355\, \mbox{\AA}$ velocity comparison for SN~2015bp and other SNe Ia. We also plot the \ion{C}{2} $\lambda 6580\, \mbox{\AA}$ and \ion{C}{1} $\lambda 10693\, \mbox{\AA}$ velocities for SN~2015bp.}
\label{fig:siIIvel}
\end{center}
\end{figure}

\begin{figure*}[htp]
\begin{center}
\includegraphics[width=1.0\textwidth]{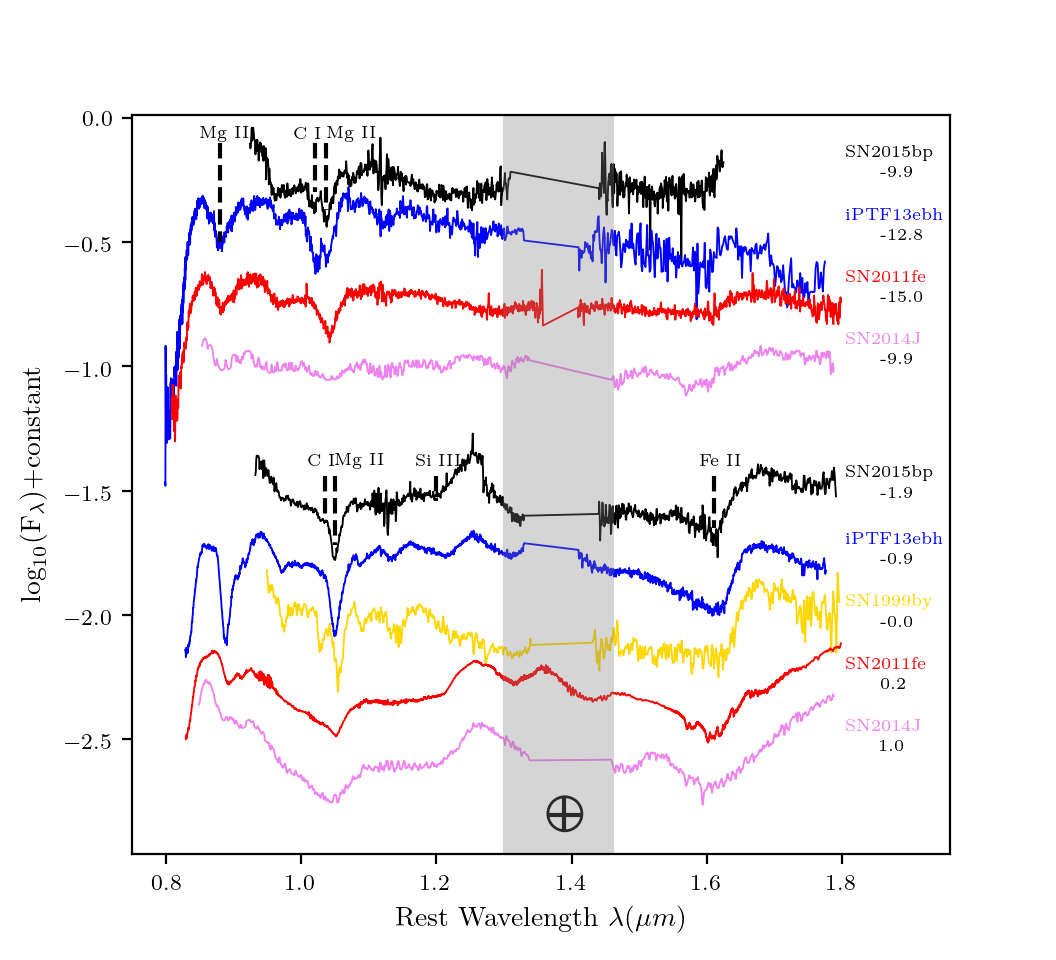}
\caption{Comparison of NIR spectra of SN~2015bp with those of other notable SNe Ia, both at a pre-maximum and near-maximum light phase. Absorption features discussed in the text are noted.
}
\label{fig:nircomp}
\end{center}
\end{figure*}

\begin{deluxetable}{ccccc}
\tablecaption{SN~2015bp velocity measurements for \ion{C}{2}  ($\lambda6580\,\mbox{\AA}$), \ion{C}{1} ($\lambda1.0693\,\mu m$), and \ion{Si}{2} ($\lambda6355\,\mbox{\AA}$). We also include the ratio of the \ion{C}{2} and \ion{Si}{2} velocities for coincident measurements. Times are measured in days relative to $B_{\rm{max}}$, and velocities are represented as $10^{3}$\,km\,s$^{-1}$. \label{tab:SiIICIIvelcomp}}
\tablehead{
    \colhead{$t_{\rm{max B}}$} & 
    \colhead{$\rm{v}_{\rm{CII}}$} & 
    \colhead{$\rm{v}_{\rm{CI}}$} & 
    \colhead{$\rm{v}_{\rm{SiII}}$} & 
    \colhead{$\rm{v}_{\rm{CII}}/\rm{v}_{\rm{SiII}}$} \\
    \colhead{$(\rm{days})$} & 
    \colhead{$10^{3}$\,km\,s$^{-1}$} & 
    \colhead{$10^{3}$\,km\,s$^{-1}$} & 
    \colhead{$10^{3}$\,km\,s$^{-1}$} & 
    \colhead{} 
}
\startdata
$-$10.9 & $13.2 \pm 0.2$ & ... & $13.0 \pm 0.1$ & $1.03 \pm 0.02$\\
$-$9.9 & ... & $11.9 \pm 0.3$ & ... & ... \\
$-$4.1 & $11.8 \pm 0.3$ & ... & $11.5 \pm 0.1$ & $1.02 \pm 0.03$\\
$-$2.8 & $11.5 \pm 0.3$ & ... & $11.3 \pm 0.1$ & $1.02\pm 0.03$\\
$-$1.9 & ... & $11.4 \pm 0.4$ & ... & ... \\
2.1 & ... & $11.1 \pm 0.5$ & ... & ... \\
\enddata
\end{deluxetable}

\subsection{Optical Properties}

As illustrated in Figure~\ref{fig:optspecevol}, SN~2015bp is clearly a SN Ia, displaying strong absorption features of ionized intermediate-mass and iron-peak elements (\ion{Si}{2}, \ion{Ca}{2}, \ion{Fe}{2}, \ion{Mg}{1}, etc), and lacking helium and hydrogen features.
In this section we go over the broad spectroscopic properties of SN~2015bp, although we will largely avoid repeating the analysis already presented by \citet{Srivastav17}, before focusing in on the prominent NIR and optical carbon features and their consequences.

In Figure~\ref{fig:spectracomp} we compare SN~2015bp with several other prominent SNe Ia at both pre-maximum and maximum light epochs\footnote{These spectra were collected from the Open Supernova Catalog \citep{opensn}: \url{https://sne.space/}}: iPTF13ebh, a transitional SN Ia \citep{iptf13ebhhsiao}; SN~2011fe, a normal SN Ia \citep{pereira2013, maguire2014}; SN~2007on, another transitional SN Ia \citep{folatelli2013, gall2018}; SN~1991bg the prototypical subluminous SN Ia \citep{filippenko1992}; and SN~1999by, another SN 1991bg-like SN Ia \citep{hoflich2002, 1999byGarn}.   Underluminous and transitional SNe Ia, in particular, have several spectroscopic features that distinguish them from normal and SN 1991bg-like events.  First, the O I $\lambda7774\,\mbox{\mbox{\AA}}$ absorption feature is more prominent in both classes of underluminous SNe Ia compared to normal objects like  SN~2011fe \citep[e.g., ][and see Figure~\ref{fig:spectracomp}]{Taubenberger08, Taubenberger17}.  The relative strength of the \ion{Si}{2} $\lambda5972\,\mbox{\AA}$ feature is greater in subluminous SNe Ia than normal objects as well, and they are generally classified as ``Cool" (CL) events in the \citet{branch2006} classification scheme.  To quantify this, we find pseudo-equivalenth width (pEW) values for \ion{Si}{2} $\lambda5972\,\mbox{\AA}$ and $\lambda6355\,\mbox{\AA}$ of $\approx35\,\mbox{\AA}$ and  $\approx108\,\mbox{\AA}$, respectively, in our spectrum taken $\sim-3$ days before the time of $B$ maximum. Other transitional SNe Ia, such as iPTF13ebh and SN~2007on, also belong to the CL subclass.   A \ion{Mg}{2} or \ion{Fe}{3} feature at $\sim$4200\,\AA\ is relatively strong in transitional SNe~Ia compared to normal SNe Ia \citep{iptf13ebhhsiao}, while a \ion{Ti}{2} ``trough" stands out in SN 1991bg-like SNe Ia at roughly the same wavelength range.  Qualitatively, SN~2015bp is a clear transitional SN Ia, displaying all of these spectroscopic hallmarks, and is remarkably similar to iPTF13ebh.

The \ion{Si}{2} $\lambda6355\,\mbox{\AA}$ absorption feature is ubiquitous in SNe Ia and is often used to measure the photospheric velocity of the SN ejecta. 
In Figure ~\ref{fig:siIIvel} we display a plot of \ion{Si}{2} velocity versus phase in comparison to other SNe Ia. The error associated with each velocity value was calculated by taking the minimum \ion{Si}{2} absorption wavelength value and calculating the velocity at $\pm2\,\mbox{\AA}$ the minimum values, and propagating it through the relativistic Doppler equation as done by \citet{silvermanvel}. 
We find that the velocity evolution of SN~2015bp follows closely that of iPTF13ebh. Since the velocity at around the time of $B_{\rm{max}}$ is $\sim$10,600\,km\,s$^{-1}$, SN~2015bp would be placed into the normal-velocity (NV) subclass as described by \cite{wang2009}.

We constrained the explosion epoch by fitting the \ion{Si}{2} $\lambda6355\,\mbox{\AA}$ velocity time evolution to a $v \propto t^{-0.22}$ power law.  As proposed by \citet{pironakar2013}, this velocity-inferred explosion time may indicate that the supernova had a ``dark phase" during the period prior to when the heating from $^{56}$Ni first reaches the outer ejecta.  This analysis implies an explosion epoch $19.08 \pm 1.68$\,days prior to  $B_{\rm{max}}$ ($JD=2,457,093.64$), where the uncertainties are inferred by calculating the explosion epoch assuming both  $v \propto t^{-0.20}$ and $v \propto t^{-0.24}$ \citep[see ][]{pironakar2013}.  This gives us a potential dark phase of 4.3\,days prior to CRTS discovery ($JD=2,457,097.9$), although since no detection limits are available, this dark phase should be considered an upper limit.  Given this, we note in passing that iPTF~13ebh had an inferred dark phase of $\sim4$\,days, so our limit is plausible.

\subsection{NIR Properties}
We also present the full NIR spectroscopic evolution of SN~2015bp in Figure \ref{fig:nirspecevol}, spanning from roughly 10 days prior through 60 days past maximum light.
We will continue our discussion of the \ion{C}{1} $\lambda1.0693\,\mu$m feature in the next section, but here we present a qualitative NIR spectroscopic comparison of SN~2015bp with the transitional SN iPTF13ebh \citep{iptf13ebhhsiao} and SN 1991bg-like SN~1999by \citep{hoflich2002}, along with the two normal and well-studied SNe Ia, SN 2011fe \citep{hsiao2013}, and SN 2014J \citep{marion2015} -- see Figure~\ref{fig:nircomp}.
One can see that at the earliest epochs, both SN~2015bp and iPTF13ebh display a strong absorption feature blueward of the \ion{Mg}{2} $\lambda\,1.0927\,\mu$m line, which we identify as \ion{C}{1} $\lambda1.0693\,\mu$m, and discuss further in Section~\ref{sec:carbcomp}.  No such feature is seen in the comparison for normal SNe Ia, and while we do not have an early-time SN 1991bg-like comparison spectrum, we do see a strong absorption feature blueward of \ion{Mg}{2} in the maximum-light spectrum of SN~1999by, which has been interpreted as \ion{C}{1} $\lambda1.0693\,\mu$m as well \citep{hoflich2002,iptf13ebhhsiao}. The strong carbon feature is weak in the transitional SNe Ia at maximum light, although there is still a faint feature in SN~2015bp which we mark in Figure~\ref{fig:nircomp}. The broad, asymmetric blue wing of the \ion{Mg}{2} $\lambda1.0927\,\mu$m absorption feature in SN~2011fe and 2014J has also been presented as evidence for \ion{C}{1} $\lambda1.0693\,\mu$m in those events \citep{hsiao2013,marion2015}.

The emission feature (and accompanying weak absorption feature blueward of it) at $\sim 1.25$\,$\mu$m has previously been identified with \ion{Si}{3} $\lambda1.2523,1.2601\,\mu$m \citep{CSPII2}.  While clearly seen in both the transitional and normal SNe Ia at maximum light, there is no similar feature in the SN 1991bg-like SN~1999by.  This absence may be real and due to the ``cool," low-temperature nature of this subclass of SNe Ia, or it may be a consequence of the lower-quality data available for SN~1999by.  Further observations of SN 1991bg-like SNe Ia in the NIR are necessary.

Another region of NIR spectroscopic interest coincides with the $H$ band where, between maximum light and $\sim+10$\,days, this region hosts a complex iron-peak emission feature that originates from allowed transitions above the photosphere \citep{wheeler1998, hoflich2002, marion2009, hsiao2013}. 
This feature can provide a constraint on the amount of $^{56}$Ni considering it consists of a blend of many \ion{Fe}{2}/\ion{Co}{2}/\ion{Ni}{2} emission lines which are produced through the radioactive decay of $^{56}$Ni. 
The study by \cite{ashall19} revealed a correlation between the light-curve shape and velocity ($v_{\rm{edge}}$) of the edge of the $H$ feature, confirming a result from \cite{hsiao2013}.
The measurement of $v_{\rm{edge}}$ corresponds to the transition in the ejecta between complete and incomplete Si-burning regions, and is determined by the exchange between the mass of $^{56}$Ni and the intermediate-mass elements formed in the explosion.
Their work included NIR spectra of SN~2015bp (labelled SNHunt281 in their study) and showed that the measurement of $v_{\rm{edge}}$ for its feature is $\sim -12,000$\,$\rm{kms}^{-1}$, which is consistent with the other transitional SNe Ia in their sample.

\section{Carbon Detections and Comparisons}\label{sec:carbcomp}

In this section we discuss the likely detection of carbon in SN~2015bp, particularly the strong \ion{C}{1} $\lambda1.0693\,\mu$m line, with respect to other SN Ia measurements in the literature.

\begin{figure*}[htp]
\begin{center}
\includegraphics[width=1.0\textwidth]{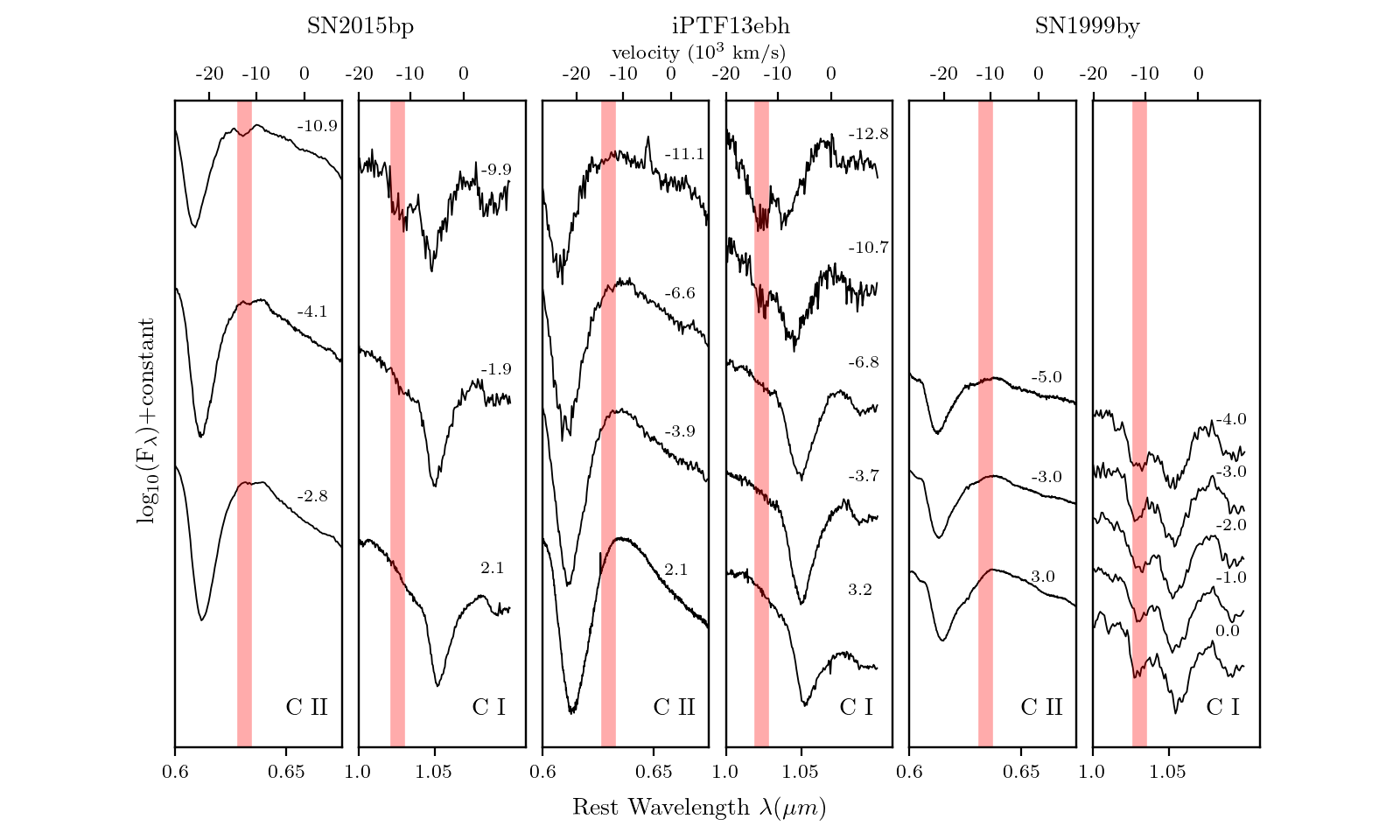}
\caption{Comparison of the optical \ion{C}{2} and NIR \ion{C}{1} evolution of
SN~2015bp with that of iPTF13ebh (another transitional SN Ia) and SN~1999by (a SN 1991bg-like SN Ia). Timescales are given in days relative to $B_{\rm{max}}$.  The red bar marks the photospheric velocity (as determined by the \ion{Si}{2} velocity) for the first epoch of observations for a given SN, and is only meant to guide the eye.  Both SN~2015bp and iPTF13ebh show early, strong \ion{C}{1} which grows weaker with time.  SN~2015bp also shows optical \ion{C}{2} absorption, while iPTF13ebh shows a ``flat" spectral feature at early times, which may also be \ion{C}{2}, but is more ambiguous.  Early-time data are not available for SN~1999by, but it shows a prominent absorption feature where one would expect \ion{C}{1}, although it remains strong through maximum light, in contrast to the transitional SNe Ia. SN~1999by also shows a flat feature where \ion{C}{2} would be expected. }
\label{fig:carboncomp13ebh}
\end{center}
\end{figure*}

We first put SN~2015bp in context with another transitional SNe Ia, iPTF13ebh, and the SN 1991bg-like SN~1999by.  We plot a time sequence around the optical \ion{C}{2} $\lambda6580\,\mbox{\AA}$ and NIR \ion{C}{1} $\lambda1.0693\,\mu$m lines for all three SNe in Figure~\ref{fig:carboncomp13ebh}.  Both SN~2015bp and iPTF13ebh show a distinct absorption feature blueward of the \ion{Mg}{2} $\lambda1.0927\,\mu$m line which grows weaker toward maximum light.  If this line is \ion{C}{1} $\lambda1.0693\,\mu$m, it would be at the photospheric velocity (see Figures~\ref{fig:siIIvel} and \ref{fig:carboncomp13ebh}). SN~2015bp also shows a clear \ion{C}{2} $\lambda6580\,\mbox{\AA}$ notch which grows weaker toward maximum light, while iPTF13ebh has a ``flat" morphology on the red shoulder of \ion{Si}{2} in the earliest spectrum, which is often recognized as a tentative \ion{C}{2} detection \citep[see, e.g.,][]{silverman2012,folatelli2012}.  Indeed, direct \texttt{SYNAPPS} modeling of iPTF13ebh indicates that this ``flat" feature is weak \ion{C}{2} \citep{iptf13ebhhsiao}.
By contrast, SN~1999by displays an apparently strong \ion{C}{1} $\lambda1.0693\,\mu$m feature up through maximum light and beyond (although no early-time data are available), staying roughly the same strength throughout.  There is a ``flat" detection of \ion{C}{2} $\lambda6580\,\mbox{\AA}$ as well, although we do not have data at early phases when \ion{C}{2} detections are more likely \citep[e.g.,][]{parrent2011}.

\begin{figure*}[htp]
\begin{center}
\includegraphics[width=0.49\textwidth]{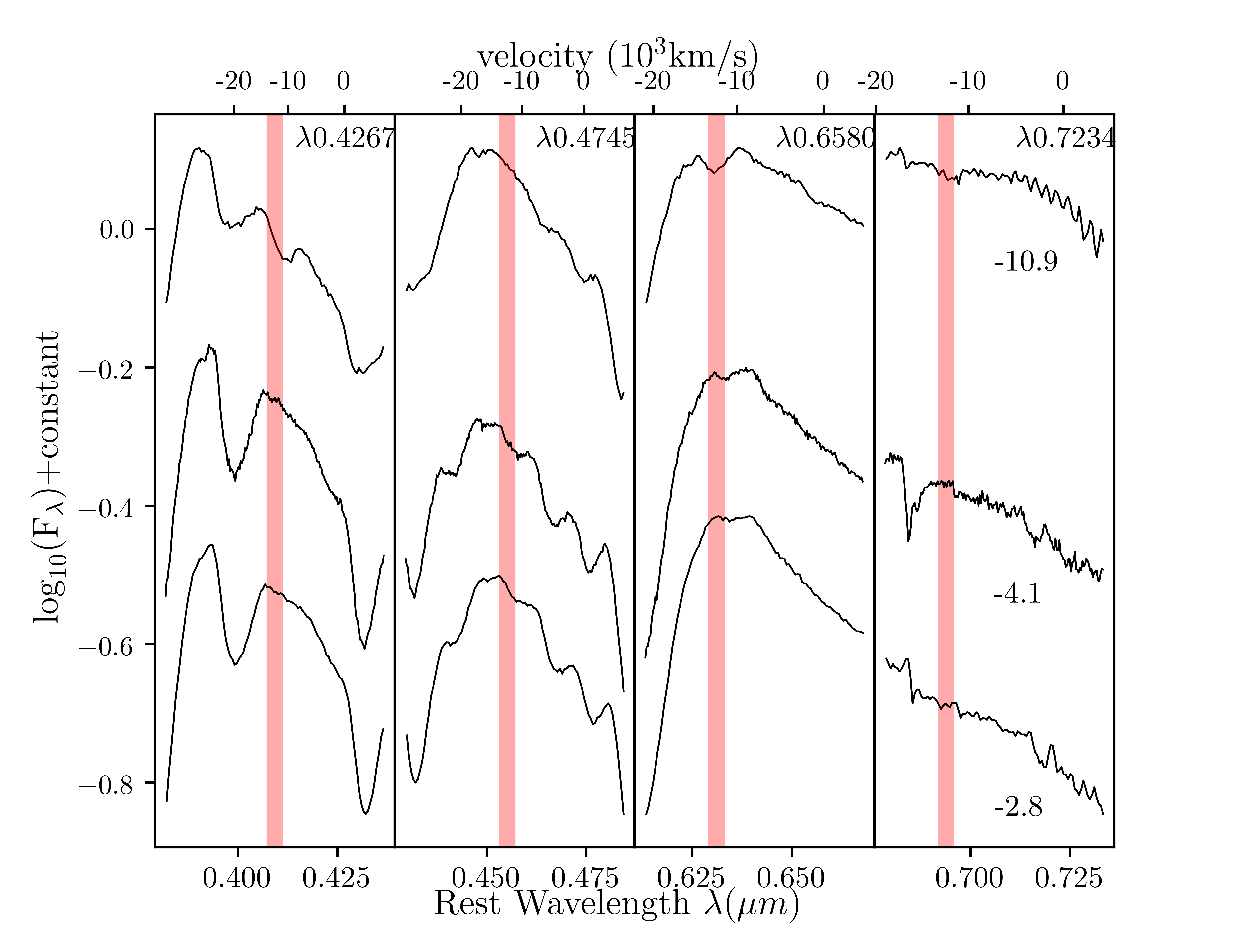}
\includegraphics[width=0.49\textwidth]{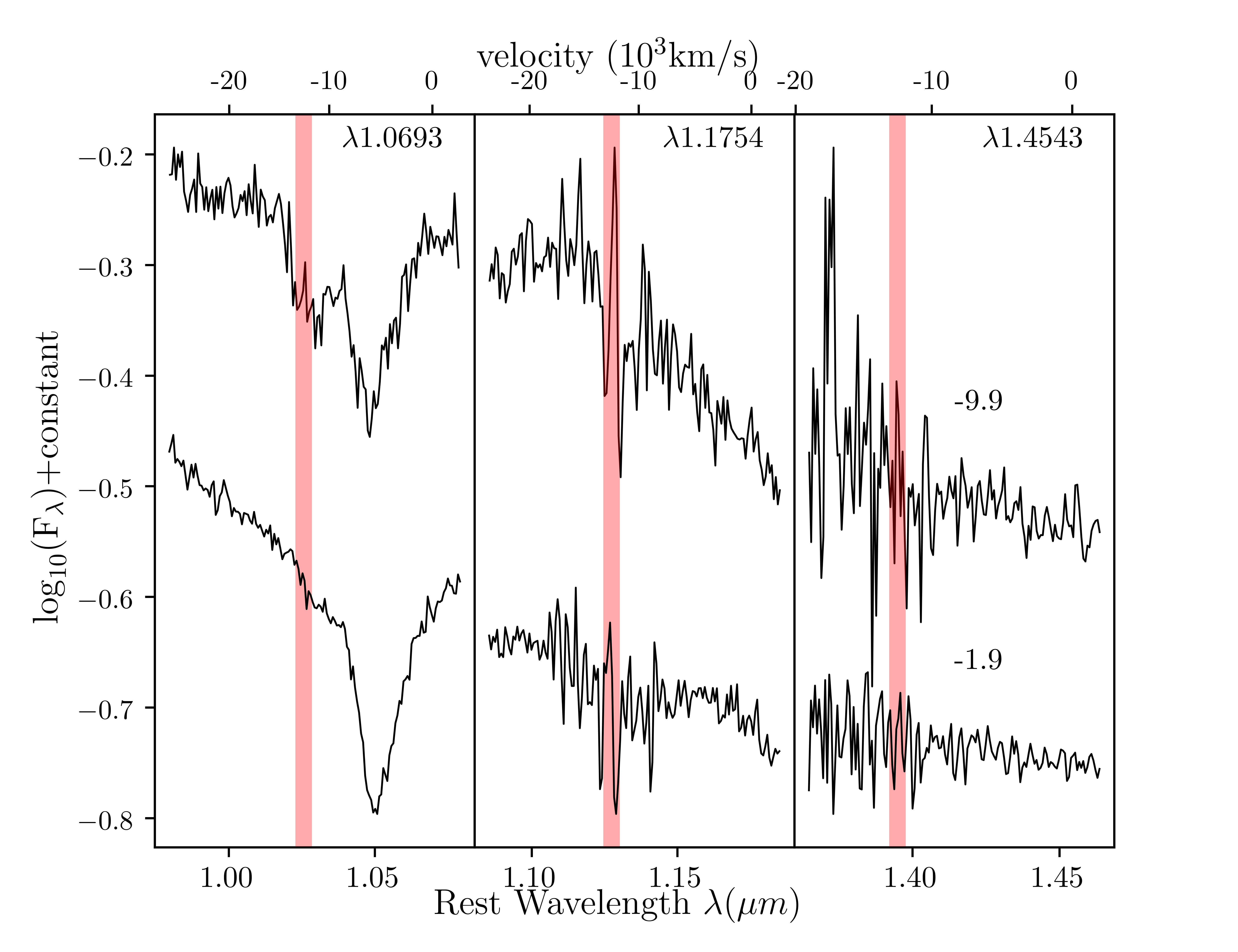}
\caption{Optical and NIR carbon lines in SN~2015bp. We highlight the region around optical \ion{C}{2} and NIR \ion{C}{1} lines beyond the standard \ion{C}{2} $\lambda6580\,\mbox{\AA}$ and NIR \ion{C}{1} $\lambda1.0693\,\mu$m features typically identified.   The panel on the left shows the three earliest optical spectra and the panel on the right displays the earliest NIR spectra (all phases with respect to $B_{\rm{max}}$). The red bar marks the photospheric velocity at $-$10.9d, and is only meant to guide the eye.  The NIR data are generally too noisy to see any features beyond \ion{C}{1} $\lambda1.0693\,\mu$m.  There may be a weak detection of \ion{C}{2} $\lambda0.7234\,\mu$m in the $-$10.9\,d optical spectrum, but otherwise no clear carbon features are apparent.
} 
\label{fig:carbonsearch}
\end{center}
\end{figure*}

In the panels of Figure \ref{fig:carbonsearch} we attempt to locate other, weak carbon features in SN~2015bp. In the optical, the only clearly visible feature is the standard $\lambda 6580\,\mbox{\AA}$ absorption line, although there is a weak notch which may be associated with \ion{C}{2} $\lambda 0.7234\,\mu$m in our $-$10.9 day spectrum.  No \ion{C}{1} absorption at $\lambda1.1754\,\mu$m or $\lambda1.4543\,\mu$m is apparent, and we note that we have effectively no data to identify the  \ion{C}{1} $\lambda0.9093\,\mu$m and $\lambda0.9406\,\mu$m features.

We investigate the extent of carbon burning by comparing the Doppler velocities of the \ion{C}{2} $\lambda6580\,\mbox{\AA}$ line with the strong \ion{Si}{2} $\lambda6355\,\mbox{\AA}$ absorption feature (see, e.g., \citealt{parrent2011}), which is typically used as a proxy for the photospheric velocity.  We display the results in Table~\ref{tab:SiIICIIvelcomp}.  The $v$(\ion{C}{2} $\lambda$6580)/$v$(\ion{Si}{2} $\lambda$6355) is near unity, with an average value of $\sim1.02$, which stays constant for the $\sim 8$ days over which we can make these measurements (from $-10.9$ to $-$2.8 days with respect to $B$ maximum).  In their much larger samples of SNe Ia, other analyses have also found roughly constant values of $v$(\ion{C}{2} $\lambda$6580)/$v$(\ion{Si}{2} $\lambda$6355), with a value slightly above unity \citep{parrent2011,silverman2012}.  This consistency amongst SNe Ia, with very few deviations, has been used as evidence that carbon is distributed in a layered (or hemispheric) geometry. A clumpy distribution will inevitably be seen at arbitrary orientations, and thus lead to some values of $v$(\ion{C}{2} $\lambda$6580)/$v$(\ion{Si}{2} $\lambda$6355)$<$1 \citep{parrent2011}.

We also measured the Doppler velocities of the NIR \ion{C}{1} $\lambda1.0693\,\mu$m line, as seen and reported in Table~\ref{tab:SiIICIIvelcomp}. Since the NIR \ion{C}{1} feature is blended heavily with the \ion{Mg}{2} $\lambda1.0927\, \mu$m absorption feature, its velocity was measured by fitting a multicomponent Gaussian and measuring the \ion{C}{1} feature at the minimum of its respective Gaussian. The velocity of the \ion{C}{1} and \ion{C}{2} features largely track each other with phase, although the earliest measurements display a $\sim 1000$\,km\,s$^{-1}$ difference, which is roughly the size of our measurement uncertainties.  This indicates that the \ion{C}{1} and \ion{C}{2} are coming from the same layer of the SN ejecta.

\begin{figure}[htp]
\begin{center}
\includegraphics[width=0.5\textwidth, height=0.7\textwidth]{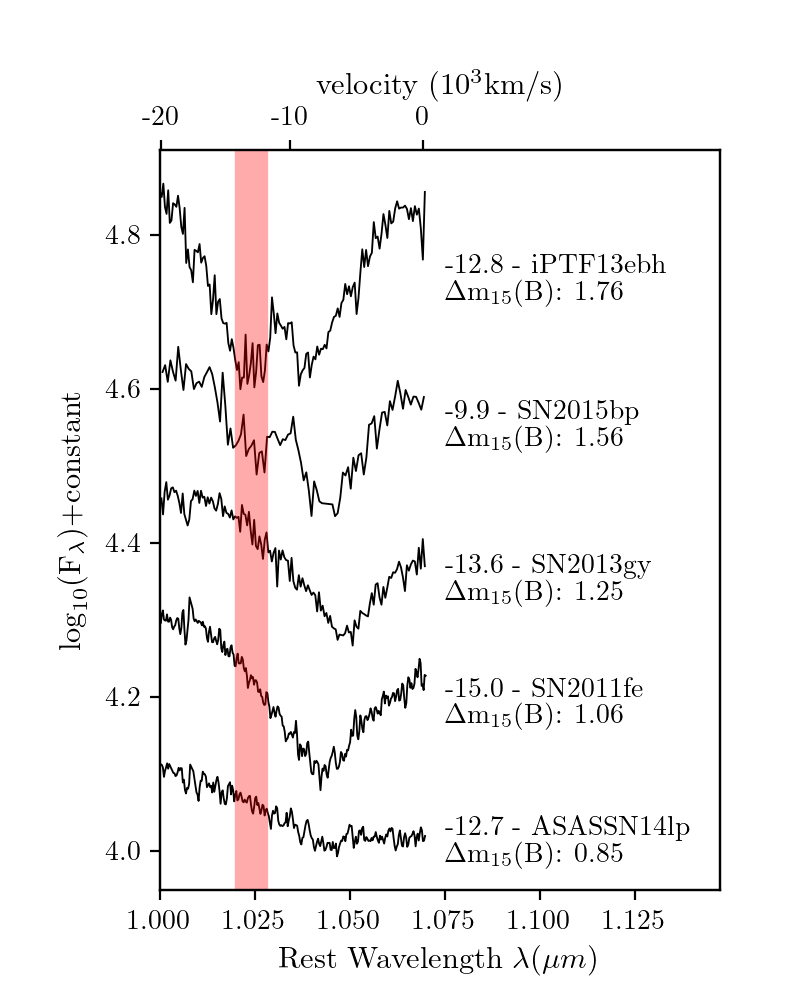}
\caption{A zoom in on the \ion{C}{1} $\lambda1.0693\,\mu$m feature for several well-studied SNe Ia with early NIR spectroscopy, plotted and ordered by $\Delta m_{15} (B)$.  Although a larger dataset is necessary, a correlation between $\Delta m_{15} (B)$ and \ion{C}{1} $\lambda1.0693\,\mu$m strength is striking. }
\label{fig:snIaearlynirCI}
\end{center}
\end{figure}

Finally, we display a zoom in on the \ion{C}{1} $\lambda1.0693\, \mu$m feature for a sample of SNe Ia with early NIR spectra ($t \lesssim -10$\,days) in Figure~\ref{fig:snIaearlynirCI}.  The sequence is plotted as a function of  $\Delta m_{15}(B)$, and shows a clear trend of stronger \ion{C}{1} for intrinsically fainter (and faster declining) SNe Ia.  In particular, transitional SNe Ia such as iPTF13ebh and SN~2015bp have distinct \ion{C}{1} absorption features at early times, while normal SNe Ia are more ambiguous.  There is an extended ``blue shoulder" apparent in the \ion{Mg}{2} line for all of the normal SNe Ia, that has been interpreted (via \texttt{SYNAPPS} modeling) as weak \ion{C}{1} \citep[see, e.g.,][]{hsiao2013,marion2015,iptf13ebhhsiao}, which if true would imply a continuous sequence of weakening NIR carbon as a function of $\Delta m_{15}(B)$ such that bright SNe Ia (slow declining) will have weaker NIR carbon.  
A similar trend has been seen in the incidence of optical \ion{C}{2}, where faster declining (intrinsically fainter) SNe Ia potentially show a clear \ion{C}{2} notch \citep[e.g.,][ although this is less conclusive in \citealt{folatelli2012}]{Thomas11,maguire2014}. A thorough investigation of transitional SN Ia will be presented in Section~\ref{sec:cii_inc}.

\section{C II Incidence in Transitional SNe Ia}\label{sec:cii_inc}

\begin{figure}
\begin{center}
\includegraphics[width=0.45\textwidth]{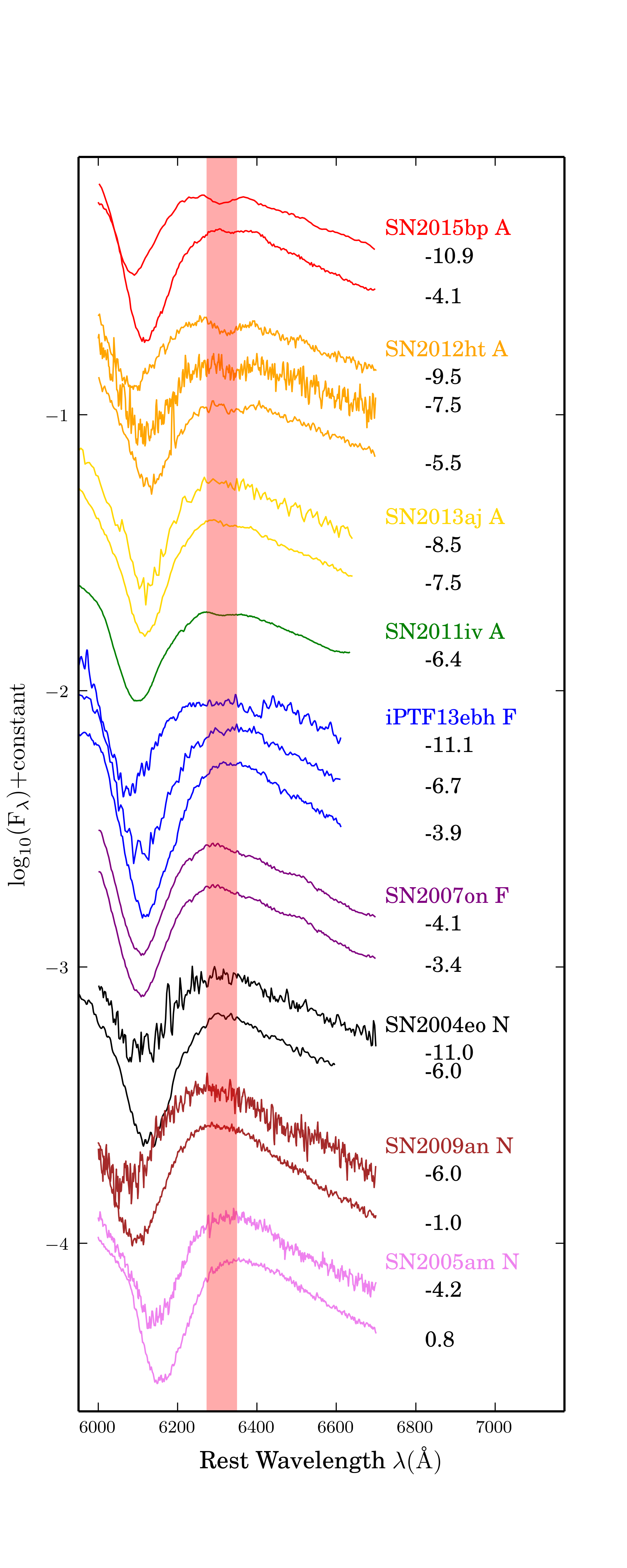}
\caption{A zoom-in view around the expected position of the \ion{C}{2} $\lambda6580\,\mbox{\AA}$ feature for our sample of nine transitional SNe Ia with early optical data.  We have classified the carbon status of each SN using the scheme of \citet{silverman2012}.  The shaded red vertical line corresponds to an expansion velocity of 13,000\,km\,s$^{-1}$ to guide the eye toward the approximate position at which carbon may be seen.}
\label{fig:trans}
\end{center}
\end{figure}

To put our carbon observations of SN~2015bp in context we have searched the literature for nearby transitional SNe Ia with early optical spectra (any data before maximum light qualified) to gauge the incidence of carbon (\ion{C}{2} $\lambda 6580 \,\mbox{\AA}$) in this subpopulation of SNe Ia.  Several programs have looked at the incidence of \ion{C}{2} $\lambda 6580\, \mbox{\AA}$ in early SNe Ia spectra \citep[e.g.,][]{Thomas11,parrent2011,folatelli2012,silverman2012,maguire2014}, but the transitional SNe Ia population itself has not been searched in a systematic way because of their relative rarity.  While we have demonstrated the possibility that transitional SNe Ia display strong \ion{C}{1} $\lambda1.0693\, \mu$m features in their early NIR spectra, most transitional SNe Ia do not have such data, and so we turn to the \ion{C}{2} $\lambda 6580\, \mbox{\AA}$ line for our analysis for further insight.  The characteristics of our collected sample are recorded in Table~\ref{tab:photoparamcomp} and the early spectra (centered on \ion{C}{2})  are shown in Figure~\ref{fig:trans}.  The specific objects are largely drawn from Table 8 of \citet{iptf13ebhhsiao}, with an additional object included from \citet{ashall2013aj}: SN~2004eo \citep{sn2004eo}, SN~2005am \citep{contreras2010, blondin2012}, SN~2007on \citep{folatelli2013}, SN~2009an \citep{sn2009an}, SN~2011iv \citep{foley2011iv, gall2018}, SN~2012ht \citep{2012ht}, SN~2013aj \citep{ashall2013aj}, and iPTF13ebh \citep{hsiao2013}.

For the nine transitional SNe Ia in our sample (including SN~2015bp), we display the 2--3 earliest available spectra in Figure~\ref{fig:trans}.  We adopt the carbon classification scheme of \cite{silvermanvel} (which is similar to most of the other large carbon analyses) where spectra with a distinct absorption feature associated with \ion{C}{2} $\lambda 6580\, \mbox{\AA}$ are distinguished by an ``A" (Absorption). Spectra that show no distinct absorption feature, but a depression or a flattening of the red side of the \ion{Si}{2} $\lambda 6355 \,\mbox{\AA}$ feature are given an ``F" (Flattened); these data represent tentative carbon detections. Finally, spectra which seem to be unaffected on the red side of \ion{Si}{2} are denoted with an ``N" (No carbon).  Our carbon classifications are included in both Table~\ref{tab:photoparamcomp} and Figure~\ref{fig:trans}.  The carbon status of several of the objects in our target sample have been remarked on elsewhere in the literature (e.g., SN~2007on, SN~2009an, SN~2012ht, and iPTF13ebh), and in all cases our carbon assessments are in agreement.  While we have not performed any spectrum synthesis modeling (e.g., \texttt{SYNOW}) to further bolster our carbon classifications, we believe this agreement with results in the literature verifies our visual assessment.

Four out of nine transitional SNe Ia in our sample display a clear \ion{C}{2} $\lambda 6580\, \mbox{\AA}$ ``notch" in our data, which we classify with an ``A" (SN~2011iv, SN~2012ht, SN~2013aj and SN~2015bp).  Two further SNe have a tentative detection with a flat, or ``F" designation (SN~2007on and iPTF13ebh).  To bolster the case that a flat designation may indeed indicate a positive carbon identification, we remind the reader that iPTF13ebh had a clear, distinct \ion{C}{1} $\lambda1.0693\, \mu$m feature in its early NIR data.  Three of our transitional SNe Ia showed no sign of carbon whatsoever (SN~2004eo, SN~2005am, and SN~2009an).  The absence of carbon in SN~2004eo is notable, as a $-$11 day spectrum is available, although it has a relatively low signal-to-noise ratio.  

The carbon statistics for our transitional SN Ia sample must be taken with caution, as previous studies have demonstrated that earlier spectra are more likely to detect \ion{C}{2} $\lambda 6580\, \mbox{\AA}$ than data taken at later times, with the \ion{C}{2} incidence going from $\sim 40$\% at $< -10$\,d to $\sim 10$--20\% in the 5 days prior to maximum light \citep{parrent2011,folatelli2012,silverman2012}.   The sample size of early transitional spectra is also small, prohibiting any strong conclusions.  With these caveats in mind, our carbon incidence rates are broadly in line with previous work on the general SN Ia population with $\gtrsim$40\% of our sample displaying a clear \ion{C}{2} notch.
Carbon is at least as common in the transitional SNe Ia as in the general population.  

With our carbon measurements of SN~2015bp and in the transitional SN Ia population in hand, we now discuss the theoretical implications for these measurements.

\section{Discussion}\label{sec:disc}

Given the consensus that SNe Ia are the thermonuclear explosion of carbon-oxygen white dwarfs, carbon provides the most direct probe of unburned material from the progenitor white dwarf, as oxygen is also produced from carbon burning.  The quantity, incidence and distribution of unburned carbon varies between SN Ia explosion models, and so observational constraints on carbon can help distinguish between models, as we have outlined in Section~1.
Here we discuss the implications of our carbon detection in SN~2015bp and the larger transitional SN Ia population.

\subsection{Sub-Chandrasekhar mass models}

Some observational studies have concluded that the faintest and fastest declining SNe Ia may result from sub-Chandrasekhar mass ejecta \citep{stritz2006,Scalzo14a,Scalzo14b}, and corresponding theoretical studies have also suggested that SN Ia on the faint end come from sub-Chandrasekhar mass explosions \citep{Blondin17,Goldstein18}.  If this is the case, and noting that most recent sub-Chandrasekhar models burn nearly all of the available carbon \citep[e.g.,][]{Fink10,polin2019}, one would expect little to no carbon in the transitional SN Ia subclass.  This is in contradiction to the clear NIR \ion{C}{1} detections in SN~2015bp and iPTF13ebh, and the prevalence of optical \ion{C}{2} we see in this population (see Section~\ref{sec:cii_inc}).

To provide further context for our transitional SN Ia \ion{C}{2} detections, we plot them in maximum light \ion{Si}{2} velocity versus $B$-band absolute magnitude space in Figure~\ref{fig:zhengpolin}.  This plot is largely a reproduction of Figure~11 in \citet{polin2019}, with the SNe Ia sample of \citet{zheng2018} plotted as black points in the background, as well as the transitional SNe Ia sample we introduced in Section~\ref{sec:cii_inc}. 

We tabulate the transitional SN Ia  data displayed in Figure~\ref{fig:zhengpolin} in Table \ref{tab:photoparamcomp}.  Absolute $B$-band magnitudes were taken from the papers listed in the final column of the table, along with the extinction applied to reach that value; we urge the reader to look directly at these references for further details, including the distances assumed in these calculations.  In most instances, we also took the peak \ion{Si}{2} velocity from the cited work, but when it was unavailable we retrieved the appropriate spectra and measured it ourselves using the techniques described in Section~4.1. As expected, the transitional SNe Ia sample sits on the faint and lower velocity end of the distribution of SNe Ia.  The dashed line shows a spline fit through several thin He shell sub-Chandrasekhar models \citep[see][for details]{polin2019} and is meant to guide the eye to show the general relationship between velocity and luminosity expected for this class of models \citep[see also][]{Shen18}.

As pointed out by \citet{polin2019}, two ``groups" or clusters of SNe Ia are apparent in Figure~\ref{fig:zhengpolin}. We note that recent work with an expanded sample indicates that these groupings may actually be part of a more continuous distribution of SN Ia properties \citep{Burrow20}; also, no host galaxy extinction is applied in the original presentation of \citet{polin2019}.  The first grouping traces the arc of SN Ia below the dashed line with the same general trend as the sub-Chandrasekhar models, while the second cluster forms a tight bunch with no apparent trend between silicon velocity and absolute magnitude. The SNe Ia in this second group are generally at lower velocities than the first group.  The suggestion of \citet{polin2019} was that these two groups represented two separate explosion mechanisms, with Chandrasekhar mass explosions related to the tightly clumped second group, while  sub-Chandrasekhar mass explosions were responsible for the first group displaying a trend between \ion{Si}{2} velocity and absolute magnitude.  

The transitional SNe Ia clearly have members that belong to both the ``Chandrasekhar" and ``sub-Chandrasekhar" subpopulations identified by \citet{polin2019}, although there is some ambiguity at the faint end of the diagram as to which population a given SN may belong.  If this inference is correct, it would suggest that multiple explosion mechanisms are responsible for transitional SNe Ia since they fall within both groups.  As can be seen in Figure~\ref{fig:zhengpolin}, there are also clear carbon detections belonging to each population.

We conclude that at least $\sim$50\% of  transitional SNe Ia in our sample do not come from sub-Chandrasekhar mass explosions due to the clear presence of carbon in their NIR and optical spectra. This statement is contingent on results from recent 1D sub-Chandrasekhar theoretical models, which show little to no carbon.  If the amount of carbon in higher dimension simulations increases (see, e.g. chapter 8 of \cite{polinthesis}), then this conclusion must be amended, but for now remains an important benchmark for theoretical interpretations of the transitional SN Ia class.
We next explore the possibility that our \ion{C}{1} detections are actually misidentified \ion{He}{1}, which may be expected in the context of double detonation models.

\subsection{Helium and carbon confusion in double detonation models}

It is possible that double detonation models, triggered by a surface detonation of helium, will leave some unburned helium behind that would be observable in the outer ejecta.  This scenario has been explored by \citet{boyle2017}, who estimated the optical depth of \ion{He}{1} lines in the ejecta of double detonation models, and computed synthetic spectra to track the purported helium line evolution with time.  
In particular, they presented calculations for the \ion{He}{1} 1.0830 and 2.0581\,$\mu$m line evolution for a ``high mass" (1.025\, $M_{\odot}$ CO core mass) and ``low mass" (0.58\, $M_{\odot}$ CO core mass) supernova model. The high-mass model is meant to correspond to a normal-luminosity SN Ia \citep{Fink10}, and the low-mass model was meant to represent underluminous, peculiar thermonuclear events \citep{Sim12}. Underluminous, transitional SNe Ia like SN~2015bp would be intermediate between these two scenarios.

In both the high-mass and low-mass models, helium absorption is visible at maximum light, and the general trend is that the helium absorption grows even stronger past maximum light (at least until +7\,d, the last epoch at which they generated spectra).  We do note that \cite{boyle2017} suggest that this evolution may be somewhat model dependent, as the helium optical depth will depend on the time evolution of the ejecta density and velocity gradient, as well as variations in the radiation temperature.  The strength of the helium absorption lines are much stronger in the low-mass model where less of the helium shell gets burned, but in both instances the   \ion{He}{1} $\lambda$1.0830\, $\mu$m feature is prominent and roughly resembles the absorption features we identified as \ion{C}{1} $\lambda1.0693\,\mu$m in SN~2015bp (as well as iPTF13ebh and SN~1999by).  Similar to our data, the \ion{He}{1} $\lambda$1.0830\, $\mu$m absorption feature is expected to sit just blueward of \ion{Mg}{2} $\lambda1.0927\,\mu$m.  Because of this similarity, \citet{boyle2017} suggested that the strong \ion{C}{1} $\lambda1.0693\,\mu$m feature identified in iPTF~13ebh and SN~1999by (and by extension, SN~2015bp) may actually be \ion{He}{1}.
However, this feature gets weaker toward maximum light in the observations of iPTF13ebh and SN~2015bp, to the point that it becomes undetectable beyond maximum light.  In contrast, the \ion{He}{1} feature seen in the \citet{boyle2017} simulations get stronger (or stays at roughly the same strength) toward maximum light and continues strengthening at later phases; the opposite of the observed trend.  Furthermore, both iPTF13ebh and SN~2015bp show optical \ion{C}{2} $\lambda6580\,\mbox{\AA}$, at least circumstantially corroborating our NIR interpretation -- these SNe Ia definitively have carbon. Given this, we find the identification of this feature with \ion{He}{1} $\lambda$1.0830\, $\mu$m to be unlikely, especially for the two transitional SNe Ia with strong features blueward of \ion{Mg}{2}.  The case of the SN 1991bg-like SN~1999by may be different, as this event had a clear absorption feature blueward of \ion{Mg}{2} that remained strong through maximum light, and did not display a clear optical \ion{C}{2} line, although the lack of early-time data precludes any definitive conclusions.  

NIR spectral time series observations of faint SNe Ia are necessary to investigate the presence of helium, and in particular high signal-to-noise ratio $K$-band spectra could enable searches for the \ion{He}{1} 2.0581\,$\mu$m line which would be less ambiguous than the \ion{He}{1} 1.0830 $\mu$m feature.  Further modeling efforts may also be required to robustly predict the appearance and time evolution of helium in double detonation models.  Nonetheless, in transitional SNe Ia the behavior of the absorption feature blueward of \ion{Mg}{2} does not match current predictions for leftover helium for double detonation SN Ia models. 

\subsection{Carbon in other scenarios}

Recent analysis of several transitional SNe Ia (SN~1986G, SN~2007on, and SN~2011iv) using radiative transfer models and the abundance stratification technique have pointed to a Chandrasekhar mass explosion consistent with delayed detonation models \citep{Ashall86G,Ashall18}, or possibly a violent merger origin \citep{Pakmor12} for SN~2007on due to its double peaked nebular emission lines \citep{Dong15, mazalli07on11iv}. This is in contrast to the studies mentioned earlier that point to a sub-Chandrasekhar origin for most faint SNe Ia \citep{Blondin17,Goldstein18}.  If transitional SNe Ia do primarily originate from Chandrasekhar mass delayed detonation explosions then we would expect to see carbon in these faint events \citep{hoflich2002}, consistent with our observations.  If some fraction originate from violent mergers, then leftover carbon would be expected as well \citep{Pakmor12}.  We find it more likely that one or both of these explosion mechanisms are responsible for the transitional SNe Ia subclass, given the observed incidence of carbon in our sample.

\begin{figure}[htp]
\begin{center}
\includegraphics[width=0.5\textwidth]{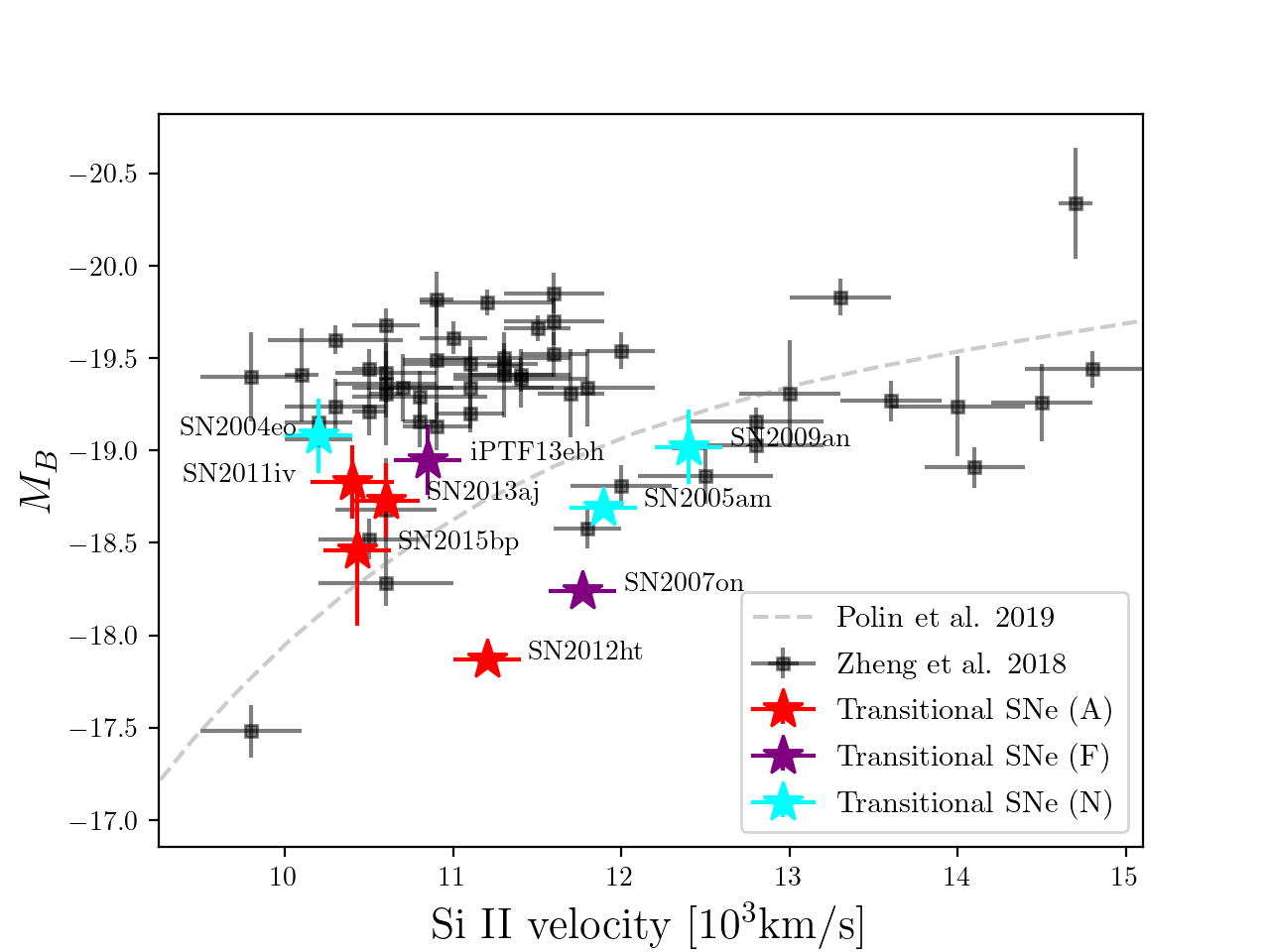}
\caption{Plot of the peak absolute magnitude in $B$  versus the \ion{Si}{2} $\lambda 6355\, \mbox{\AA}$ velocity at the time of peak for the \cite{zheng2018} collection of SNe Ia. We also plot the spline fit taken from the thin helium shell ($0.01 M_{\odot}$) sub-Chandrasekhar double detonation models in \cite{polin2019}. We have included our sample of transitional SNe Ia (see Table~\ref{tab:photoparamcomp}) with early optical spectra into this plot as well, marking those with and without optical carbon in their early data (A and N, respectively).  We have also labeled those objects with a `flat' profile, which we consider a tentative \ion{C}{2} detection (F).
}
\label{fig:zhengpolin}
\end{center}
\end{figure}

\begin{deluxetable*}{ccccccccc}
\tablecaption{Carbon Classification and Photometric Parameters of Transitional SNe Ia \label{tab:photoparamcomp}}
\tablehead{\colhead{SN} & \colhead{$\Delta m_{15}(B)$} & \colhead{$s_{bv}$} & \colhead{Earliest Spec.} & \colhead{\ion{C}{2} $\lambda6580\,\mbox{\AA}$} & \colhead{$M_{\rm{B}}$} & \colhead{\ion{Si}{2}} & \colhead{$E(B-V)^c$} & \colhead{References} \\
\colhead{} & \colhead{(mag)} & \colhead{} & \colhead{Phase (days)$^b$} & \colhead{class$^a$} & \colhead{(mag)} & \colhead{$10^3$~km/s} & \colhead{} & \colhead{\textit{see Notes}}}
\startdata
SN~2004eo & $1.42 \pm 0.03$ & $0.83 \pm 0.03$ & $-11.0$ & N & $-$19.08 & 10.2 & 0.093 & (d) \\ 
SN~2005am & $1.53 \pm 0.02$ & $0.75 \pm 0.02$ & $-$4.2  & N & $-$18.69 & 11.9 & 0.043 & (e, f)\\
SN~2007on & $1.96 \pm 0.01$ & $0.57 \pm 0.04$ & $-$4.1  & F & $-$18.24 & 11.8 & 0.001 & (g)\\
SN~2009an & $1.64 \pm 0.04$ & $0.86 \pm 0.06$ & $-$6.0  & N & $-$19.02 & 12.4 & 0.108  & (h,m)\\
SN~2011iv & $1.77 \pm 0.01$ & $0.64 \pm 0.04$ & $-$6.4  & A & $-$18.83 & 10.4 & 0.001 & (i, j) \\
SN~2012ht & $1.30 \pm 0.04$ & $0.86 \pm 0.03$ & $-$9.5  & A & $-$17.87 & 11.2 & 0.025 & (k)\\
SN~2013aj & $1.47 \pm 0.01$ & $0.78 \pm 0.01$ & $-$7.5  & A & $-$18.73 & 10.6 & 0.032 & (l)\\
iPTF13ebh & $1.79 \pm 0.01$ & $0.63 \pm 0.02$ & $-$11.1& F & $-$18.95 & 10.9 & 0.116 & (m) \\
SN~2015bp & $1.56 \pm 0.03$ & $0.671 \pm 0.030$ & $-$10.9& A & $-$18.46 & 10.4 & 0.047 & This work\\
\enddata
\tablecomments{$^a$The carbon classification scheme is adopted from \cite{silvermanvel} where an ``A" denotes a clear absorption feature, ``F" indicates a depression or flattening of the red side of the \ion{Si}{2} $\lambda 6355\, \mbox{\AA}$, and ``N" denotes spectra that do not display any carbon absorption.}
\tablecomments{$^b$ The phase is with respect to the time of $B$-band maximum}
\tablecomments{$^c$ Total extinction applied, including Milky Way and host components.}
\tablecomments{$^d$ \cite{sn2004eo}, $^e$ \cite{contreras2010}, $^f$ \cite{blondin2012}, $^g$ \cite{folatelli2013}, $^h$ \cite{sn2009an}, $^i$ \cite{foley2011iv} $^j$ \cite{gall2018}, $^k$ \cite{2012ht}, $^l$ \cite{ashall2013aj}, $^m$ \cite{iptf13ebhhsiao}}
\end{deluxetable*}

\section{Summary \& Conclusion}\label{sec:conc}

We have presented comprehensive observations of the transitional SN Ia 2015bp, the highlight of which was the early, striking detection of \ion{C}{1} $\lambda1.0693\,\mu$m.  We placed this carbon detection in context with other early NIR observations, and  assessed the incidence of carbon in the transitional SN Ia population more generally.    A summary of  our main findings is as follows.

\begin{itemize}
\item SN~2015bp displays all of the trademarks of the transitional SN Ia class \citep[see, e.g.,][]{iptf13ebhhsiao}.  It is subluminous ($M_B=-18.46$\,mag) with a fast-declining light curve ($\Delta m_{15}(B)=1.56$\,mag; $s_{BV} = 0.67$), and a NIR primary maximum that occurs before $B$-band maximum.  No \ion{Ti}{2} absorption is apparent in the optical spectra.  The light curve and spectroscopic parameters we measured for SN~2015bp are largely in agreement with previous work on this object \citep{Srivastav17}.

\item Early-time NIR spectra of SN~2015bp exhibit a prominent absorption line in the blue wing of \ion{Mg}{2} $\lambda1.0927\,\mu$m which we attribute to \ion{C}{1} $\lambda1.0693\,\mu$m.  This feature weakens in strength through maximum light; similar to that seen in transitional SN Ia iPTF13ebh \citep{iptf13ebhhsiao}. The SN 1991bg-like SN~1999by, by contrast, displayed a strong \ion{C}{1} $\lambda1.0693\,\mu$m feature past maximum light \citep{hoflich2002}.

\item In addition to the \ion{C}{1} $\lambda1.0693\,\mu$m line, SN~2015bp also displays a clear \ion{C}{2} $\lambda$6580\,\AA\ notch at early times.  The velocity of the \ion{C}{1} and \ion{C}{2}  features are consistent with each other to within $\sim 1000$\,km\,s$^{-1}$, and are marginally above the photosphere as measured by the \ion{Si}{2} $\lambda$6355\,\AA\ absorption line velocity.

\item There appears to be a correlation between the strength of the \ion{C}{1} $\lambda 1.0693\, \mu$m absorption feature and the light-curve decline rate in SNe Ia with NIR data at very early times ($\lesssim -10$\,d; see Figure~\ref{fig:snIaearlynirCI}) in the sense that faster-declining events have more apparent carbon \citep[see also][]{hoflich2017, CSPII2}.   Of note, it is only in fast-declining SNe Ia that the \ion{C}{1} $\lambda 1.0693\, \mu$m feature is conspicuous in the blue wing of \ion{Mg}{2}.  Further data of a variety of SNe Ia at early times are needed to verify this and to confirm the trend mentioned above. 

\item Given the presence of strong NIR \ion{C}{1} in the early-time spectra of both SN~2015bp and iPTF13ebh, we investigated the incidence of early optical  \ion{C}{2} $\lambda$ 6580\,\AA\ in a larger sample of transitional SNe Ia.  Four out of nine transitional SNe Ia in our sample display a clear \ion{C}{2} notch, while two others have tentative ``flat" detections.  With the caveat that we are in the small-numbers regime, our carbon incidence rate is broadly in line with previous work on the general SN Ia population, and stands at $\gtrsim$40\%.

\item We find it unlikely that the NIR \ion{C}{1} feature is actually misidentified \ion{He}{1}, as suggested for double-detonation models \citep{boyle2017}.  The observed \ion{C}{1} feature seen in SN~2015bp and iPTF13ebh weakens up through maximum light, while the \ion{He}{1} predicted by the double-detonation models strengthens over the same time period.  The clear presence of optical \ion{C}{2} in SN~2015bp further bolsters the case that the strong NIR absorption feature is \ion{C}{1}.

\item The presence of strong NIR carbon in SN~2015bp and iPTF13ebh, along with the incidence of optical carbon in the transitional SNe Ia class, argues against a sub-Chandrasekhar origin for these faint SNe Ia despite recent modeling efforts \citep{Blondin17,Goldstein18}. Carbon in the transitional SNe Ia is consistent with originating from Chandrasekhar-mass delayed detonation and/or violent merger explosions, as other work has suggested \citep[e.g.,][]{Ashall86G,Ashall18}.

%ORIGINAL: \item The presence of strong NIR carbon in SN~2015bp and iPTF13ebh, along with the incidence of optical carbon in the transitional SNe Ia class as a whole, argues against a sub-Chandrasekhar origin for these faint SNe Ia despite recent modeling efforts \citep{Blondin17,Goldstein18}.  Carbon is consistent with transitional SNe Ia originating from Chandrasekhar-mass delayed detonation and/or violent merger explosions, as other work has suggested \citep[e.g.,][]{Ashall86G,Ashall18}.

\end{itemize}

Several avenues of future research present themselves; while we focus on work associated with the detection of carbon here, other areas such as nebular spectroscopy of comprehensive samples of faint type Ia SNe may also prove fruitful.  First, very early-time NIR spectroscopic sequences of Type Ia SNe are still rare, but it is clear that more datasets are necessary to solidify tentative correlations between NIR \ion{C}{1} strength and light-curve decline rate, as suggested by Figure~\ref{fig:snIaearlynirCI}.  Extending these early-time observations to include high signal-to-noise ratio $K$-band observations would allow useful searches for helium, a prediction of double-detonation models. Furthermore, larger samples of early-time optical spectra may identify the incidence of carbon among subpopulations of Type Ia SNe (i.e., as a function of position on the iconic Branch diagram or the ``Polin plot" as seen in Figure~\ref{fig:zhengpolin}) and would  go hand in hand with modeling efforts as the community tries to understand the viable explosion mechanisms for SNe Ia. Indeed, a couple of recent, faint type I supernovae -- SN~2018byg \citep{De19} and SN~2016jhr \citep{Jiang17} -- do have the hallmarks of a sub-Chandrasekhar double detonation explosion, showing that this explosion model is viable, at least for peculiar SNe. This comprehensive study of SN~2015bp and the transitional SNe Ia population in general show the continued promise of unburned carbon for testing explosion models.

\acknowledgments
We thank D. K. Sahu for providing early-time optical spectra of SN~2009an.

Research by D.J.S. is supported by NSF grants AST-1821967, AST-1821987, AST-1813708, AST-1813466, and AST-1908972, as well as by the Heising-Simons Foundation under grant \#2020-1864.  The CSP-II has been supported by National Science Foundation (NSF) grants AST-1008343, AST-1613426, AST-1613455, and AST-1613472, as well as by the Danish Agency for Science and Technology and Innovation through a Sapere Aude Level 2 grant. E.Y.H. and J.L. also acknowledge the support of the Florida Space Grant Consortium.
This work was partially performed at the Aspen Center for Physics, which is supported by NSF grant PHY-1607611. 
Research by S.V. is supported by NSF grants AST–1813176 and AST-2008108.
L.G. was funded by the European Union's Horizon 2020 research and innovation programme under the Marie Sk\l{}odowska-Curie grant agreement No. 839090. This work has been partially supported by the Spanish grant PGC2018-095317-B-C21 within the European Funds for Regional Development (FEDER).
M.S. is supported by generous grants from Villum FONDEN (13261, 28021) and by a project grant (8021-00170B) from the Independent Research Fund Denmark. 
A.V.F. is grateful for financial assistance from the TABASGO Foundation, the Christopher R. Redlich Fund, and the Miller Institute for Basic Research in Science (U.C. Berkeley).

Based on observations obtained at the international Gemini Observatory (GN-2015A-Q-8, GS-2015A-Q-5), a program of NSF’s NOIRLab, which is managed by the Association of Universities for Research in Astronomy (AURA) under a cooperative agreement with the National Science Foundation. on behalf of the Gemini Observatory partnership: the National Science Foundation (United States), National Research Council (Canada), Agencia Nacional de Investigaci\'{o}n y Desarrollo (Chile), Ministerio de Ciencia, Tecnolog\'{i}a e Innovaci\'{o}n (Argentina), Minist\'{e}rio da Ci\^{e}ncia, Tecnologia, Inova\c{c}\~{o}es e Comunica\c{c}\~{o}es (Brazil), and Korea Astronomy and Space Science Institute (Republic of Korea).  This paper includes data gathered with the  Nordic Optical Telescope (PI Stritzinger) at the Observatorio del Roque de los Muchachos, La Palma, Spain.
This work is based in part on observations from the DEep Imaging Multi-Object Spectrograph at the Keck-2 telescope. We are grateful to the staff at the Keck Observatory for their assistance, and we extend special thanks to those of Hawaiian ancestry on whose sacred mountain we are privileged to be guests. The W. M. Keck Observatory is operated as a scientific partnership among the California Institute of Technology, the University of California, and NASA; it was made possible by the generous financial support of the W. M. Keck Foundation. We thank S. Bradley Cenko for  assistance with the Keck spectral reductions, as well as Patrick Kelly, WeiKang Zheng, and John Mauerhan for their assistance with the observations.

D.J.S is a visiting Astronomer at the Infrared Telescope Facility, which is operated by the University of Hawaii under contract 80HQTR19D0030 with the National Aeronautics and Space Administration.  Based	on	data	products	from	observations	made	with	ESO	Telescopes	at	the	La	Silla
Paranal	Observatory	under	programmes 188.D-3003 and	191.D-0935:		PESSTO (the Public	ESO	Spectroscopic Survey for Transient Objects).	

\facilities{Swope, Du Pont (Retrocam), NTT (EFOSC2, SOFI), NOT (ALFOSC), Gemini:Gillett (GNIRS), Gemini:South (F2), IRTF (SpeX), Magellan:Baade (FIRE)}

\software{Astropy \citep{astropy}, IDL \citep{IDLastro}, Matplotlib \citep{Matplotlib}, NumPy \citep{numpy}, \texttt {SNooPy}  \citep{Burns11}}

\bibliographystyle{aasjournal}
\bibliography{SN15bp}

\end{document}